\documentclass[twocolumn,showpacs,preprintnumbers,amsmath,amssymb,pre]{revtex4}

\usepackage[dvips]{graphicx}
\usepackage{dcolumn}
\usepackage{amssymb}
\usepackage{bm}

\begin{document}

\preprint{APS}

\title{Granular contact force density of states and entropy in a modified Edwards ensemble}

\author{Philip T. Metzger}
\email{Philip.T.Metzger@nasa.gov}
\affiliation{%
The KSC Applied Physics Laboratory, John F. Kennedy Space Center, NASA\\
YA-C3-E, Kennedy Space Center, Florida  32899
}%

\date{\today}

\begin{abstract}
A method has been found to analyze Edwards' granular contact force probability functional for a special case.  As a result, the granular contact force probability density functions (PDFs) are obtained from first principles for this case.  The results are in excellent agreement with the experimental and simulation data.  The derivation assumes Edwards' flat measure---a density of states (DOS) that is uniform within the metastable regions of phase space.  The enabling assumption, supported by physical arguments and empirical evidence, is that correlating information is not significantly recursive through loops in the packing.  Maximizing a state-counting entropy results in a transport equation that can be solved numerically.  For the present this has been done using the ``Mean Structure Approximation,'' projecting the DOS across all angular coordinates to more clearly identify its predominant non-uniformities.  These features are: (1) the Grain Factor $\overline{\Psi}$ related to grain stability and strong correlation between the contact forces on the same grain, and (2) the Structure Factor $\overline{\Upsilon}$ related to Newton's third law and strong correlation between neighboring grains.
\end{abstract}

%\keywords{Suggested keywords TBD}
\maketitle

\section{\label{sec:intro}Introduction}

\subsubsection{Deriving the Contact Force Distribution}

\begin{figure}
\includegraphics[angle=-90,width=0.45\textwidth]{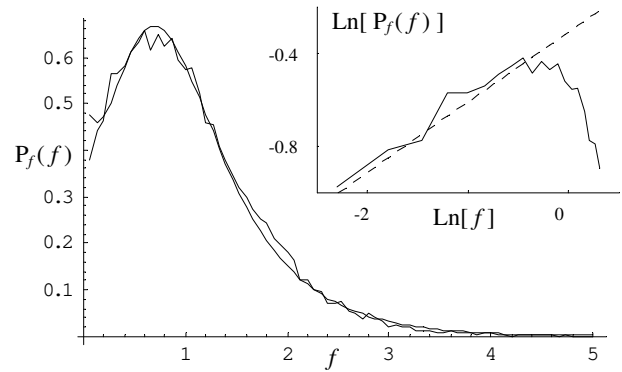}
\caption{\label{fig:vectorPDF} Linear plot of the PDF $P_{f}\left(f\right)$ of the normalized vector magnitudes of the granular contact forces resulting from Monte Carlo solution of the Mean Structure Transport Equation.  It has a non-zero probability density for zero force, a peak just below $f=1$, and an exponential tail with decay constant $\beta = 1.6$.  The smooth curve is a fit to Eq.~(\ref{eqn:chicagofit}).  The log-log inset shows the behavior below $f=1$.   The dashed line is a power law of exponent $\alpha = 0.3$.  These features are consistent with experimental and simulation data.}
\end{figure}

There have been several attempts to derive the granular contact force probability density function (PDF) for static granular packings, $P_{F}(F)$, by using analogies from thermal statistical mechanics \cite{bagi, bagi2, kruytroth, ngan, edwards2}.  The interest arises in part because the empirical $P_{F}(F)$ \cite{silbertgrest, landry, erikson, chicago, ohern1, snoeih1, snoeih2, frictionless, radjaijean, radjai, thornton, antony} has an exponential tail, reminiscent of the energy distributions of thermal systems.  However, the overall form of $P_{F}(F)$ is not found in thermal systems, generally having a peak or plateau near the average force and a non-zero value at zero force as illustrated in Fig.~(\ref{fig:vectorPDF}).

In contrast to this form, the prototypical distributions found in thermal systems are either monotonically decreasing (e.g., the Gibbs energy distribution), or begin from zero probability density at the origin before rising to a peak (e.g., the Maxwell-Boltzmann distribution).  In the non-monotonic cases the rising slope is due to the degeneracy of energy states.  The degeneracies reflect the dimensionality of the system and dominate the form of the distribution at weak energies beginning from the origin.  Since the forces in a granular medium are vector magnitudes composed from several Cartesian components---implying degeneracy in the force magnitudes---this raises the question why $P_{F}(F)$ does not likewise begin from the origin $P_{F}(0)=0$ before rising to its peak?  Indeed, a recent model \cite{edwards2, brujic} predicts that it should.  The model represents a first-principles approximation for key elements of the physics and results in a Boltzmann-type equation that is solvable.  This produces a $P_F(F)$ that begins from $P_F(0)=0$, rises to a peak, and then decays exponentially.  Because of these considerations, the question may be asked whether the empirical observations that $P_F(0)>0$ is primarily the result of numerical or experimental uncertainties:  the distribution is in question precisely where the forces are weakest and therefore most difficult to model or measure.  Perhaps the theory provides a clearer view into the fundamental organization of the Density of States (DOS) in this region than the empirical methods are presently able to provide.  

It seems to the author that this is not the case for two reasons.  First, it has been shown that the form in the region of weak forces evolves in a predictable way as a function of stress and/or fabric anisotropy, which may be induced through shearing \cite{antony}.  The anisotropy dependence probably explains the variations in $P_F(F)$ seen among the different empirical studies, in that some jammed packings have displayed peaks while others have displayed plateaus or monotonic forms.  For a packing of grains originally in an isotropic state, $P_{F}(F)$ displays a form similar to Fig.~(\ref{fig:vectorPDF}).  As the packing is quasi-statically sheared the distribution smoothly evolves to having a plateau in the region of weak forces and then on to becoming a monotonically decreasing function with only an abrupt change of slope where the peak had previously been.  After the packing achieves peak shear strength, continued shearing reduces the stress anisotropy and causes the distribution to retrace its evolution most of the way, ending with a small peak again.  This behavior affects the distribution well above the region of numerical uncertainty and cannot be dismissed as the result of dynamical or transient forces since the shearing is quasi-static.  It is difficult to see how this smooth variation of forms---including plateaus and monotonic forms---could be explained if the finite $P_{F}(0)>0$ were not real.  

Second, the unique features of the PDF have been obtained using a wide range of empirical techniques, and it does not seem reasonable that all of them are incorrect in the region of weak forces.  These techniques include experiments with frictional grains \cite{chicago, erikson}, numerical simulations with frictional grains \cite{landry,silbertgrest, radjaijean, radjai, antony} or purely frictionless grains \cite{snoeih1, frictionless, snoeih2}, and adaptive network models \cite{tkachenko2}.  The simulation techniques have included contact dynamics (CD), discrete element modeling (DEM), and molecular dynamics (MD) quenched beneath the glass transition \cite{ohern1}, all of which are well-established techniques.  The contact laws in these simulations have included Hertzian, Hookean, and Lennard Jones potentials.  Simulations have been done with and without gravity and under a wide variety of conditions.  The transitions between the boundary and bulk have been thoroughly examined \cite{snoeih2}.  The numerical techniques have demonstrated the ability to distinguish between distributions that begin at the origin and those that do not \cite{snoeih2}.  Although experiments with frictionless emulsions \cite{brujic} and some numerical simulations \cite{thornton} have been fitted to forms that begin with $P_{F}(0)=0$, arguably those data would be fit as well or better by forms with nonzero $P_{F}(0)$.

The universality of these observations shows that the PDF's unique features are not associated with a specific type of model or the (non)existence of friction, but are fundamental characteristics of static granular packings.  Because of this, the present paper will proceed with the assumption that these observations are correct but have yet to be explained.  Perhaps the explanation lies in a unique generalization of statistical mechanics.  Just as the DOS for ideal Bose and Fermi gases are organized differently than the classical dilute gas and therefore produce their own unique energy distributions, so the DOS of granular packings may be organized in some unique way to produce this distinctive PDF.

Such a generalization has been taking shape \cite{theory, EdwStress1, bouchaud}, beginning with Edwards' hypothesis \cite{edwards1} that all metastable packings are equally probable in the statistical ensemble.  Another line of progress is based on the concept of directed force chain networks \cite{directedchain}, while others aim to understand the distribution of forces beneath a localized perturbation or more generally the stress response function \cite{response}, and the phenomena related to jamming and unjamming \cite{jamming, ohern3}.  This paper focuses more narrowly upon those models or hypotheses which predict a PDF by making assumptions about the DOS in the ensemble, including those models which take a random walk in a phase space (e.g., the $q$ model) or a PDF space (i.e., the Boltzmann transport equation variety), and those which directly assume the form of an entropy or other thermodynamic functional.

The $q$ model \cite{liu, coppersmith} may be considered a random walk because the set of forces in a single layer of the lattice describe a locus in phase space while the random redistribution of those forces from one layer to the next (controlled by the stochastic $q$ variables) represents a random walk through that space.  Eventually the walk wanders into regions of the space having the most probable distribution of coordinates.  Bouchaud has shown that the sufficient requirement to obtain the exponential tail in the $q$ model is simply that some grains transmit all their load from one hemisphere into just one contact on the other hemisphere \cite{bouchaud}.  This introduced a new way to think about granular media:  the statistical relaxation of the force distribution does not occur dynamically through the time dimension as it does in thermal systems; rather it is a necessary feature of the \textit{internal}, layer-by-layer static equilibrium relationships, where the spatial dimensions play a role analogous to the time dimension for the corresponding set of Cartesian components of force \cite{coppersmith}.  Several generalizations of the $q$ model and other lattice-based models have been developed \cite{lattice}.  Some of these are similarly random walks in a non-dynamic phase space, but others include explicitly dynamic features to recursively achieve organization in the percolating force network.

In this context it is probably helpful to mention again \cite{radjainote} that the distribution predicted by the $q$ model \cite{coppersmith} was not $P_F(F)$, but rather $P_w(w)$ where $w$ is the total vertical load supported by the grain. Distributions of $w$ and $F$ have been occasionally confused with one another, especially since $w$ and $F$ become identical in the special case at the flat sides of a container.  This has contributed to the confusion over the form of $P_F(F)$.  The $q$ model can also produce distributions $P_{X}(F_{x})$ of the vertical Cartesian components of the contact forces, $F_{x}$, but it cannot directly predict the vector magnitudes $F$ of those same contact forces.  The $P_{X}(F_{x})$ predicted by the $q$ model is always monotonically decreasing, in agreement with numerical simulation data \cite{snoeih2}.

Another theoretical model that makes direct statements about the contact force DOS is the Boltzmann-type equation presented by Edwards and Grinev \cite{edwards2, brujic} mentioned above.  In the discussion section, this paper shall attempt to reconcile the model with the empirical data.  

Other models include several entropy maximization or functional minimization concepts.  These methods produce elements of the empirically-observed PDFs, but not all of their features.  The concept proposed by Bagi \cite{bagi, bagi2} deals, like the $q$ model, with Cartesian components.  It produces the same Canonical distribution as the uniform $q$ model.  The concept proposed by Kruyt and Rothenburg \cite{kruytroth} deals with contact force magnitudes and predicts $P_{F}(0) = 0$, a peak, and an exponential tail.  The concept proposed by Ngan \cite{ngan} produces $P_{F}(0) > 0$, a peak, and a nearly-Gaussian, compressed-exponential tail.  Unlike Edwards and Grinev's model, these last three are not derived from first-principles but are hypotheses drawn by analogy with other entropic systems.  Despite any shortcomings, all these models provide important insights into the nature of the PDF problem.

\subsubsection{Organization of the Paper}

This paper is organized as follows.  Sec.~II will present a first-principles analysis of the DOS in a modified version of the Edwards ensemble \cite{edwards1}.  The dynamical behaviors of granular media will be completely avoided so that Edwards' hypothesis alone shapes the DOS.  It will be shown that the DOS is highly self-organized and very sparse.  Its form depends upon the form of $P_{F}(F)$ and \textit{vice versa} so that recursion is necessary to solve for either.  Maximizing a state counting entropy in this phase space will produce the recursion equation which is analogous to the Boltzmann transport equation.  To elucidate the organization of the DOS, it will be projected in the Mean Structure Approximation across all angular coordinates before solving numerically.  Sec.~III presents the numerical solution of the Mean Structure Transport Equation.  The results demonstrate the success of Edwards' hypothesis in that it predicts a form for the PDFs in qualitative and quantitative agreement with the experimental and simulation data, having $P_{F}(0) > 0$, a peak, and an exponential tail with a decay constant matching empirical observations.  Sec.~IV discusses the validity of the approach and insights into the physics that produce the features of $P_{F}(F)$.  Sec.~V summarizes the paper and points to several unanswered problems and generalizations that are needed.

\section{\label{sec:analysis}Modified Edwards Ensemble Analysis}

\subsection{Description of the Particular Ensemble}

Following Edwards and coworkers, this analysis focuses upon 2D, amorphous packings of cohesionless, rigid grains having the fixed coordination number that makes the packing isostatic \cite{EdwStress1}.  The problem shall be further idealized, however, by using only smooth, round grains.  Also, this paper focuses on the frictionless case wherein the 2D isostatic coordination number is $Z=4$.  A method has been found to solve Edwards' probability functional for this special case.  Although this ensemble is highly idealized, it is a good starting point because 2D packings of cohesionless, round grains that are perfectly rigid \cite{radjaijean, radjai} and/or frictionless \cite{ohern1, ohern3, frictionless, snoeih2, silbertgrest, brujic} are known to have force distributions with the same qualitative features as the more generalized packings and hence must be subject to the same organizational constraints in the statistics.  Therefore they are sufficient to elucidate the origin of those constraints in the physics.

The use of exactly four contacts per grain, however, is more idealized than has been achieved in typical numerical simulations.  Nevertheless, it is acceptable because that is the average coordination number for 2D frictionless packings of round, rigid grains which are isostatic \cite{bouchaud, isostatic}, and it will be shown herein that the same qualitative and quantitative features of $P_{F}(F)$ arise as they do in the more realistic simulations.

Two defining issues for the ensemble are (1) how to specify the fabric, and (2) how to apply stress to it.  Since this paper is concerned primarily with the derivation of $P_{F}(F)$, and since its form is known to evolve with stress and fabric anisotropy under shearing, the ensemble will sufficiently general to accommodate anisotropy in each.  On the other hand, this paper does not address the more ambitious problem of stress propagation.  Therefore the analysis shall not accommodate large-scale stress and fabric inhomogeneities that persist in the ensemble average.  An example of an inhomogeneous case is the conical sandpile formed by central pouring, which has directional fabric and stress propagation away from the center of the pile \cite{sandpile}.  In this paper, only actions such as shaking, shearing and compressing are assumed to have occurred in the construction history because these produce homogeneous stress and fabric states with only statistical fluctuations, vanishing in the ensemble average.

Specifically, stress will be specified as the tensor which is a volume average over the entire packing.  The source of stress will be mechanical at the boundaries to allow for the full range of possible states, something which gravity alone cannot do.  Gravity will be eliminated both because it is not necessary and because it breaks a symmetry of the ensemble and may thus tend to obscure the organizational features of the physics.  

This leaves the question how to specify the fabric.  Edwards and coworkers have developed a conjugate field theory with the goal of explaining the propagation of stresses in granular materials correlated to the local contact geometry \cite{EdwStress1}.  In that theory it has been shown that two fabric tensors are required to produce the complete set of stress propogation equations.  They have elsewhere developed a thermodynamic theory of compaction, in which the relevant specifier is simply the scalar volume of the packing \cite{edwards1}.  For the sake of simplicity, the choice was made there to avoid the full anisotropic treatment.  For the present purposes, something is needed which is less than the full tensorial treatment but more than scalar compaction.  We will therefore use the joint probability density function (JPDF) $P_{4\theta}(\theta_{1},\ldots,\theta_{4})$ studied in Ref.~\cite{troadec}.  This function correlates the contact angles that share the same grain. It can be collapsed to $P_{\theta}(\theta)$,
\begin{equation}
P_{\theta}(\theta) = \frac{1}{4}\sum_{\beta=1}^{4}\int\!\!\!\int\!\!\!\int\!\!\!\int_{0}^{2 \pi}\!\!\!\!\text{d}^{4}\theta \ P_{4\theta}(\theta_{1},\ldots,\theta_{4})\ \delta(\theta-\theta_{\beta}).\label{collapse}
\end{equation}
The use of the uncollapsed distribution is deemed necessary because the physics of fragile media are grain-centered and the intra-grain correlations turn out to be the heart of the statistical physics, as shall be shown here.  This JPDF is not sufficient to relate the packing state to the specified stress tensor, however, because it tells nothing of the number and size of voids created by the arrangement of neighboring grain configurations.  For the present, the voids may be quantified simply by assuming a number of grains per unit distance in a cross-section of the 2D packing in each orthogonal direction.  These quantities along with the JPDF will be specified in the ensemble rather than predicted.  Evolution of the internal state of the packing is beyond the present study.

Finally, for convenience the idea of ``quartered fabric'' is introduced at several points.  It is defined such that $P_{4\theta}(\theta_{1},\ldots,\theta_{4})$ is zero everywhere except where the $j^{\text{th}}$ contact is on the $j^{\text{th}}$ quadrant of every grain in the packing.  For the specific case of ``quartered isotropy,'' collapsing the quartered fabric by Eq.~(\ref{collapse}) produces $P_{\theta}(\theta)=1/2\pi$.  This mimics true isotropy but the anisotropic quartering is apparent in the JPDF.  As in the case of non-quartered fabric, $P_{4\theta}$ enforces steric exclusion.  To achieve quartered isotropy with steric exclusion in a Monte Carlo process it is necessary to weight the distribution of attempted angles to emphasize the regions close to the edges of each quadrant.  Otherwise, steric exclusion would cause notches to appear in $P_{\theta}(\theta)$ near the boundary of each quadrant.  The use of quartered fabric in this analysis is only to provide insight into the expressions.  It is always possible to write and numerically solve them for the more general case, and it was found that numerical solutions were indistinguishable with or without quartered fabric.

\subsection{The Phase Space}

The locus in phase space of a classical dilute monatomic gas completely defines its state.  We wish to define a phase space for granular packings which is similarly complete.  A 2D frictionless granular packing of $N$ round, rigid grains is isostatic and therefore contains $2N$ contacts.  The phase space therefore requires at least $4N$ phase space axes, half of which define the force on each contact and half of which define the contact angles, $\{F_{k},\theta_{k}\mid k=1,\ldots,2N\}$, which is labeled $\mathbb{S}_{1}$ and has DOS $\rho^{(1)}$.  It is possible to define the ordering of the axes so that it is understood which four contacts correspond to the same grain and therefore which grains contact one another.  It will not prove necessary to do so explicitly, although this ordering is implicitly assumed to exist.  

Newton's third law (N3L) is automatically satisfied in $\mathbb{S}_{1}$, since each contact is represented by only one force and one angle axis.  However, enforcing Newton's Second Law (N2L) will prove simpler if redundant axes are created to account for each contact force twice, one time with each grain that shares the contact, $\{F_{\alpha \beta},\theta_{\alpha \beta}\mid \alpha=1,\ldots,N; \beta=1,\ldots,4\}$, where $\alpha$ subscripts the grain and $\beta$ subscripts the contact on the grain.  This space is labeled $\mathbb{S}_{2}$.  In this new space it will be necessary to enforce N3L.  Again, it is possible to define the ordering of the axes so that it is understood which contacts are redundant to one another and therefore which grains are contacting neighbors.  It will not be necessary to do so explicitly, although this ordering is implicitly assumed to exist.

In the thermodynamic limit $N\to\infty$ this ensemble has Edwards' flat measure, every metastable state being equally probable,
The DOS in $\mathbb{S}_{2}$ is,
\begin{eqnarray}
 \rho^{(2)}\{F_{\alpha \beta},\theta_{\alpha \beta}\} & = & \delta(\text{fabric } P_{4\theta})\nonumber\\*
& \times & \delta\!\left(\sum_{\alpha} w_{x \alpha}- W_{x} \right) \delta\!\left(\sum_{\alpha} w_{y \alpha}- W_{y} \right)\nonumber\\*
& \times & \left\{\prod_{\text{contacts}} \delta\!\left(\vec F_{\gamma \delta}+\vec F_{\epsilon \zeta}\right)
\right\}\nonumber\\*
& \times &  \prod_{\alpha=1}^N \delta\!\!\left(\sum_{\beta=1}^{4} \vec F_{\alpha \beta}\right) \ \prod_{\beta=1}^{4} \Theta\left( F_{\alpha \beta} \right)
\label{DOS2}
\end{eqnarray}
where $\Theta$ is the Heaviside (unit step) function.  

The six constraints which define the accessible regions of phase space are described below.  

1.  The JPDF for the fabric is specified by the first delta function.  Actually, there should be a statement relating the continuum $P_{4\theta}(\theta_{1},\theta_{2},\theta_{3},\theta_{4})$ with the discretized distribution of angles at finite $N$, $\mu_{klmn}(\theta_{1k},\ldots,\theta_{4n})$, but the meaning is nonetheless clear.

2. (and 3.)  The Cartesian loads $w_{x}$ and $w_{y}$ on each grain will often be called the ``supported loads'' or simply the ``loads''.  At each locus in phase space the relationship exists between these loads and the Cauchy stress tensor.  (See for example Ref.~\cite{cauchy}.)  For simplicity the sum over these Cartesian loads is specified.  Hence,
\begin{equation}
\sum_{\alpha=1}^{N} w_{x \alpha}= W_{x},\ \ \ \sum_{\alpha=1}^{N} w_{y \alpha}= W_{y},
\end{equation}
where the loads are defined by,
\begin{equation}
\begin{array}{ll}
w_{x \alpha}=\left(w_{x \alpha}^{\text{left}}+w_{x \alpha}^{\text{right}}\right)/2,\ \ \ &
w_{y \alpha}=\left(w_{y \alpha}^{\text{top}}+w_{y \alpha}^{\text{bottom}}\right)/2,
\end{array}
\end{equation}
and, using non-quartered fabric,
\begin{equation}
\begin{array}{lll}
w_{x \alpha}^{\text{left}} & = & -\sum_{\beta=1}^{4}L_{\alpha \beta}F_{\alpha \beta}\cos\theta_{\alpha \beta},\\
w_{x \alpha}^{\text{right}} & = & +\sum_{\beta=1}^{4}R_{\alpha \beta}F_{\alpha \beta}\cos\theta_{\alpha \beta},\\ 
w_{y \alpha}^{\text{top}} & = & +\sum_{\beta=1}^{4}T_{\alpha \beta}F_{\alpha \beta}\sin\theta_{\alpha \beta},\\
w_{y \alpha}^{\text{bott.}} & = & -\sum_{\beta=1}^{4}B_{\alpha \beta}F_{\alpha \beta}\sin\theta_{\alpha \beta}.
\end{array}
\label{wmapping}
\end{equation}
The operator $L_{\alpha \beta}$ multiplies the expression by 1 if $\pi/2\le\theta_{\alpha \beta}<3\pi/2$ meaning the $\beta^{\text{th}}$ contact is on the left half of the grain, else it multiplies it by 0.  Likewise $R_{\alpha \beta}$, $T_{\alpha \beta}$ and $B_{\alpha \beta}$ test for contacts on the right, top and bottom side of the grain, respectively.  For stable grains, $w_{x \alpha}=w_{x \alpha}^{\text{left}}=w_{x \alpha}^{\text{right}}$, but the hemispheric distinctions shall be useful in the analysis.

4.  N3L is satisfied between every pair of contacting grains.
\begin{equation}
\vec F_{\gamma \delta}=-\vec F_{\epsilon \zeta},
\end{equation}
where grains $\gamma$ and $\epsilon$ are contacting neighbors through their $\delta^{\text{th}}$ and $\zeta^{\text{th}}$ contacts, respectively.

5.  N2L for static equilibrium must be satisfied at each grain individually,
\begin{equation}
\sum_{\beta} \vec F_{\alpha \beta}=0\ \forall\ \alpha.
\end{equation}

6.  $\Theta$ enforces no tensile contacts anywhere in the packing, which restricts the DOS to the first ``quadrant'' of the force axes.

In addition to these six constraints, two missing constraints should be noted:

1.  The above ensemble does not enforce the shear stress but relies on the fact that their ensemble average is zero and in the thermodynamic limit the fraction of packings in which the shear deviates from zero by more than some arbitrarily small amount will vanish.  The Cartesian axes of the packings are taken to be aligned with the principle stress axes so that the off-diagonal elements of the stress tensor should be zero.  

2.  A cluster of real grains must just touch one another, forming closed loops, but in the above ensemble the geometric constraints for grains outside the first coordination shell have been intentionally omitted.  This \textit{First Shell Approximation} (FSA) asserts that only negligible correlative information travels all the way around closed loops of grains in the ensemble average.  In other words, the DOS is adequately characterized for the present purposes by the two-point (intra-grain) force correlations and the resulting correlation of loads in neighboring grains.  Therefore, the geometric closure of force loops can be ignored when deriving the statistics of single-grain states.  There are important arguments supporting the FSA and they will be presented in the discussion section.

\subsection{Phase Space Operations to Quantify the Non-Uniformity}

Although Edwards' flat measure is uniform across the regions of accessible phase space, the volume of those regions is not uniformly distributed across the coordinates.  The program is to change coordinates in a way that eliminates delta function constraints from the right-hand side of Eq.~(\ref{DOS2}), trading the lack of uniformity in the constraints for a lack of flatness in the measure.  When only extensive, conserved quantities remain in the list of constraints, then the method of the most probable distribution may be used, relying on the method of Lagrange multipliers to conserve those quantities.

\subsubsection{Newton's Third Law}

In this context the term ``grain configuration'' refers not only to a grain's contact geometry but also to the set of forces upon those contacts as defined by the locus in phase space.  The form of $\mathbb{S}_{2}$ itself does not require neighboring grains to satisfy N3L, and so the vast majority of loci include neighboring grain configurations with physically unrealizable forces.  This mathematical abstraction enables the analysis.  

If we wished to neglect N3L then we could proceed with the remainder of the analysis, obtaining the hypothetical DOS for regions of this space having stable, cohesionless grains, write the state-counting entropy and then maximize it subject to the conservation of total loads and fabric.  This would produce the most probable distribution for all possible permutations of $N$ stable grain configurations where the grains are mechanically \textit{independent}.  However, it turns out that N3L is not negligible:  by considering instead all the possible \textit{combinations} (rather than all possible \textit{permutations}) of $N$ stable, independent grain configurations, we note that some of these combinations can be mechanically connected into a greater number permutations satisfying N3L than can other combinations.  Hence, those combinations are the more entropic ones, the ones which represent the greater number of metastable packings in the phase space.  Therefore, to find the most entropic combination of stable, independent grain configurations, we will map $\mathbb{S}_{2} \to \mathbb{S}_{3}$, a space where the axes are the same as in $\mathbb{S}_{2}$ except that they are not sequenced to represent a particular permutation of the grains.  Whereas a locus in $\mathbb{S}_{2}$ represents a single state (a single packing permutation of a set of grain configurations), a locus in $\mathbb{S}_{3}$ represents a set of mechanically disconnected grain configurations that may or may not be permutable into some number of stable states.  We shall call the latter an ``assembly space'' to distinguish it from a phase space that identifies every grain's location in the packing.  The fraction of permutations that satisfy N3L will be quantified in this mapping process.

To verify that a particular permutation of a given combination of stable grains $\{F_{\alpha \beta},\theta_{\alpha \beta}\}$ satisfies N3L, it is necessary to check every contact in the permutation.  All permutations of this combination have the same JPDF of forces and contact angles, $P_{F\theta}(F,\theta)=P_{F\theta}(F,\theta \mid \{F_{\alpha \beta},\theta_{\alpha \beta}\})$, whatever its form may be.  Randomly choosing one contact from the set of these permutations, a contact force $F_{\alpha \beta}$ at angle $\theta_{\alpha \beta}$ therefore has the probability $P_{F\theta}(F_{\alpha \beta},\theta_{\alpha \beta})\text{d}F\text{d}\theta$ that it will satisfy N3L with its neighbor.  (The two differentials reflect the fact that N3L reduces the solution space by two dimensions per contact, thereby taking out the extra dimensions introduced in $\mathbb{S}_{1} \to \mathbb{S}_{2}$.)  The probability that an entire grain configuration drawn from this set of permutations will satisfy N3L with its four neighbors will be called $\Upsilon^{2}(F_{\alpha 1},\ldots,F_{\alpha 4},\theta_{\alpha 1},\ldots,\theta_{\alpha 4})\ \text{d}^{4}F\ \text{d}^{4}\theta$, written for compactness as $\Upsilon^{2}(F_{\alpha \beta},\theta_{\alpha \beta})\ \text{d}^{4}F\ \text{d}^{4}\theta$.  It may be written as a functional of $P_{F\theta}$,
\begin{eqnarray}
\Upsilon^{2}(F_{\alpha \beta},\theta_{\alpha \beta}) & = & \prod_{\beta} P_{F\theta}\left(F_{\alpha \beta},\theta_{\alpha \beta}\right).
\label{Ups1}
\end{eqnarray}
This expression treats the contacts on the neighboring grains as if they are uncorrelated because this is a packing that was drawn randomly from the set of all possible permutations, including the ones which are physically unrealizable.  Therefore there are no \textit{a priori} correlations between neighboring grains; such correlations arise \textit{a posteriori} by selecting the subset of packings that satisfy N3L.  

Because of this statistical independence, the fraction of packings that satisfy N3L for all of its grains is likewise simply the product over the probabilities that each of the individual grains will satisfy N3L with its own local neighbors.  (The FSA appears implicitly in this statement.)  However, the product of $\Upsilon^{2}(F^{\alpha \beta},\theta^{\alpha \beta})$ over all $\alpha$ accounts for every contact in the packing twice, once with each grain sharing the contact.  For the cases where N3L is in fact satisfied, the double accounting of contacts will appear as pairs of $P_{F\theta}$ factors having identical arguments.  Hence, the probability that the entire packing will satisfy N3L is the square root of that product---explaining the use of the square exponent in Eq.~(\ref{Ups1}).  The fraction of permutations that satisfy N3L is,
\begin{equation}
\Phi_{\text{N3L}}\{F_{\alpha \beta},\theta_{\alpha \beta}\} = \prod_{\alpha=1}^{N}\Upsilon(F^{\alpha \beta},\theta^{\alpha \beta})\ \text{d}^{2N}\!F\ \text{d}^{2N}\!\theta. \label{PN3L}
\end{equation}
This calculation does not handle the boundaries of the packing (unless they are periodic), but we are concerned with the statistics in the bulk in the thermodynamic limit where the boundaries are pushed out toward infinity.  

Now the DOS may be mapped from the phase space to the assembly space, $\mathbb{S}_{2} \to \mathbb{S}_{3}$,
\begin{eqnarray}
\tilde{\rho}^{(3)}\{F_{\alpha \beta},\theta_{\alpha \beta}\} = \delta(\text{fabric } P_{4\theta})\ \delta\!\left(\sum_{\alpha} w_{x \alpha}- W_{x} \right)\nonumber\\*
\times\ \delta\!\left(\sum_{\alpha} w_{y \alpha}- W_{y} \right)\ \prod_{\alpha=1}^N \Upsilon(F_{\alpha \beta},\theta _{\alpha \beta})\ \text{d}^{2N}\!F\ \text{d}^{2N}\!\theta\nonumber\\*
\times\ \delta\!\!\left(\sum_{\beta=1}^{4} \vec F_{\alpha \beta}\right)\ \prod_{\beta=1}^{4} \Theta\left( F_{\alpha \beta} \right)\nonumber\\*\ \label{DOS3}
\end{eqnarray}
The tilde on $\tilde{\rho}$ indicates that this density is in an assembly space.

\subsubsection{Newton's Second Law}

To quantify the effects of N2L, note that Eq.~(\ref{wmapping}) can be used as a many-to-one mapping from $\mathbb{S}_{3} \to \mathbb{S}_{4}$, which will have coordinates $\{w_{\xi \alpha},\theta_{\alpha \beta}\mid \xi=x,y; \alpha=1,\ldots,N; \beta=1,\ldots,4\}$ and is another assembly space, representing combinations of mechanically-independent grain configurations.  Thus, the mapping reduces the dimensionality of the space by two per grain, just as N2L reduces the dimensionality of the solution space by two per grain.  However, the reverse mapping is one-to-one because the localized isostacy of the grains determines the four contact forces when the supported loads and four contact angles are specified.  Thus, of all the points in $\mathbb{S}_{3}$ that map to the same point in $\mathbb{S}_{4}$, at most only one represents a stable packing and is occupied, the one which is specified by solving Eq.~(\ref{wmapping}) for $F_{\alpha \beta}$ with $w_{x}=w_{x}^{\text{left}}=w_{x}^{\text{right}}$ and $w_{y}=w_{y}^{\text{top}}=w_{x}^{\text{bott.}}$.  This system of equations is nonsingular except for some precise alignments of contacts on a grain which we can ignore.  The Jacobian of transformation for $\mathbb{S}_{3} \to \mathbb{S}_{4}$ is simple to write and is a functional of the fabric.  Instead, $\Upsilon$ is re-defined to produce the Jacobian with delta functions,
\begin{eqnarray}
\Upsilon^{2}(F_{\alpha \beta},\theta_{\alpha \beta}) & = & \int\!\!\!\int\!\!\!\int\!\!\!\int_{o}^{\infty}\!\!\!\!\!\text{d}^{4}F \prod_{\beta} P_{F\theta}\left(F_{\alpha \beta},\theta_{\alpha \beta}\right) \nonumber\\*
& & \times\ \delta\left(w_{x \alpha}-F_{\alpha 1}\cos\theta_{\alpha 1}-F_{\alpha 4}\cos\theta_{\alpha 4}\right)\nonumber\\*
& & \times\ \delta\left(w_{x \alpha}+F_{\alpha 2}\cos\theta_{\alpha 2}+F_{\alpha 3}\cos\theta_{\alpha 3}\right)\nonumber\\*
& & \times\ \delta\left(w_{y \alpha}-F_{\alpha 1}\sin\theta_{\alpha 1}-F_{\alpha 2}\sin\theta_{\alpha 2}\right)\nonumber\\*
& & \times\ \delta\left(w_{y \alpha}+F_{\alpha 3}\sin\theta_{\alpha 3}+F_{\alpha 4}\sin\theta_{\alpha 4}\right).\nonumber\\*
\ \label{ups4}
\end{eqnarray}
(Note that for simplicity this has been written with quartered fabric.)  With this, the DOS in $\mathbb{S}_{4}$ may be written,
\begin{eqnarray}
 \tilde{\rho}^{(4)}\{w_{\xi \alpha}, \theta_{\alpha \beta}\} & = & \delta(\text{fabric } P_{4\theta})\nonumber\\*
 & \times & \delta\!\left(\sum_{\alpha} w_{x \alpha}- W_{x} \right)\delta\!\left(\sum_{\alpha} w_{y \alpha}- W_{y} \right)\nonumber\\*
 & \times & \prod_{\alpha=1}^N \Upsilon(w_{x \alpha},w_{y \alpha},\theta_{\alpha \beta})\ \text{d}^{2N}\!\theta\nonumber\\*
& \times & \prod_{\beta=1}^{4} \Theta\left[ F_{\alpha \beta}(w_{x \alpha},w_{y \alpha},\theta_{\alpha \beta}) \right].\nonumber\\*
\label{DOS4}
\end{eqnarray}

\subsubsection{Cohesionless Grains}

The product over Helmholtz functions may simply be omitted if it is remembered that the $P_{F \theta}(F,\theta)$ factors contained within $\Upsilon$ must be zero for negative arguments $F$.  On the other hand, it will prove convenient to define $\Psi(F_{\alpha \beta},\theta_{\alpha \beta})=\prod_{\beta=1}^{4} \Theta(F_{\alpha \beta})$ where $F_{\alpha \beta}=F_{\alpha \beta}(w_{x \alpha},w_{y \alpha},\theta_{\alpha \beta})$.

\subsection{State-Counting Entropy and Its Maximum}

Randomly drawing packings from the regions of $\mathbb{S}_{4}$ that have a specified $P_{2w4\theta}(w_{x},w_{y},\theta_{1},\ldots,\theta_{4})$ (and hence a specified fabric and a specified $P_{F\theta}(F,\theta)$), the fraction of packings in which all grains will satisfy N2L without cohesion \textit{and} satisfy N3L with its neighbors is,
\begin{equation}
\begin{array}{l}
\Phi\{w_{\xi \alpha},\theta_{\alpha \beta}\}\left[P_{F\theta}\right] = \\*
\ \\*
\begin{array}{lll}\ \ & = & \prod_{\alpha} \Upsilon(F_{\alpha \beta},\theta_{\alpha \beta})\left[P_{F\theta}\right]\cdot \Psi(F_{\alpha \beta},\theta_{\alpha \beta})\ \text{d}^{2N}\theta \\*
\ \\*
& = & \prod_{i}\cdots \prod_{n} \Big[\Upsilon(w_{xi}, w_{yj}, \theta_{1k},\ldots,\theta_{4n})\left[P_{F\theta}\right]\\*
& & \ \ \times \Psi(w_{xi}, w_{yj}, \theta_{1k},\ldots,\theta_{4n}) \Big]^{\nu_{ijklmn}}\ \text{d}^{2N}\theta\label{PhiDist1}
\end{array}
\end{array}
\end{equation}
where $\nu_{ijklmn}(w_{xi}, w_{yj}, \theta_{1k},\ldots,\theta_{4n})$ is the discretized version of the distribution $P_{2w4\theta}(w_{x},w_{y},\theta_{1},\ldots,\theta_{4})$, normalized such that $\sum_{i}\cdots \sum_{n} \nu_{ijklmn} = N$.  It was obtained by discretizing the $(w_{x}, w_{y}, \theta_{1}, \ldots, \theta_{4})$ space into bins of volume $(\Delta w_{x} \cdot \Delta w_{y} \cdot \Delta\theta_{1} \cdots \Delta\theta_{4}) = (\Delta w)^{2}(\Delta \theta)^{4}$.  Note that in Eq.~(\ref{PhiDist1}), the product in the first line is over the grains whereas the products in the second line are over the discretized intervals of each of the variables $w_{x}$, $w_{y}$, $\theta_{1}, \ldots, \theta_{4}$.  For compactness we will write,
\begin{equation}
\Phi\{w_{\xi \alpha},\theta_{\alpha \beta}\}\left[P_{F\theta}\right] = \prod_{i}\!\cdots\!\prod_{n}\!\Big[\Upsilon_{i \ldots n} \cdot \Psi_{i \ldots n}\Big]^{\nu_{i\ldots n}}\!\text{d}^{2N}\theta.\label{PhiDist2}
\end{equation}

To find the most probable $P_{2w4\theta}(w_{x},w_{y},\theta_{1},\ldots,\theta_{4})$ that results from the non-uniform DOS of Eq.~(\ref{DOS4}), we likewise discretize $\mathbb{S}_{4}$ into cells of volume $(\delta w_{x}\cdot\delta w_{y}\cdot \delta\theta_{1}\cdots\delta\theta_{4})^{N}$ where $(\delta w)^{2}(\delta\theta)^{4} = (\Delta w)^{2}(\Delta\theta)^{4}/S$, where $S$ is a large integer and $S>>\nu_{i\ldots n} \forall\ (i,\ldots,n)$.  The number of cells in $\mathbb{S}_{4}$ which map to a particular set $\{\nu_{i\ldots n}\}$ can be estimated by explicit counting,
\begin{eqnarray}
\omega\{\nu_{i\ldots n}\}& = & \prod_{i}\cdots \prod_{n}\left[\frac{(S-1+\nu_{i\ldots n})!}{(S-1)!\ (\nu_{i\ldots n})!}\right]\nonumber\\*
& & \times\left(\sum_{i}\cdots \sum_{n}\nu_{i \ldots n}\right)!\label{locuscount}
\end{eqnarray}
and in the limit as $S \to \infty$ the estimate becomes exact.  However, because $\mathbb{S}_{4}$ is an assembly space, the axes can be relabeled $N!$ different ways to represent the same combination of grains.  Removing this physically meaningless repetition, we omit the factorial of the sums in the second line of Eq.~(\ref{locuscount}),
\begin{equation}
\tilde{\omega}\{\nu_{i\ldots n}\} = \prod_{i}\cdots \prod_{n}\left[\frac{(S-1+\nu_{i\ldots n})!}{(S-1)!\ (\nu_{i\ldots n})!}\right]\label{combocount},
\end{equation}
where the tilde on $\tilde{\omega}$ indicates that this is the ``correct Boltzmann counting'' for an assembly space, in which the grains are indistinguishable.

The number of states $\Omega$ in the ensemble mapping to the distribution $\{\nu_{i\ldots n}\}$ is therefore $\tilde{\omega}\{\nu_{i \ldots n}\}$ times the DOS in those cells of $\mathbb{S}_{4}$,
\begin{eqnarray}
\Omega\{\nu_{i\ldots n}\} & = & \prod_{i}\cdots \prod_{n}\left[\frac{(S-1+\nu_{i\ldots n})!}{(S-1)!\ (\nu_{i \ldots n})!}\right] \nonumber\\*
& & \times\ N!\ \left[\Upsilon_{i\ldots n}\Psi_{i\ldots n}\right]^{\nu_{i\ldots n}}\ \text{d}^{2N}\theta
\end{eqnarray}
where we have used the notation of Eq.~(\ref{PhiDist2}) to express the DOS.  Taking the logarithm, it may be maximized with respect to $\nu_{p\ldots u}$.  If we discretize the JPDF of fabric $P_{4\theta}(\theta_{1},\ldots,\theta_{4}) \to \mu_{klmn}(\theta_{k},\ldots,\theta_{n})$ such that $\sum_{k}\cdots\sum_{n} \mu_{klmn}=N$, then each value of $\mu_{klmn} \forall k,l,m,n$ is a conserved quantity according to the definition of the ensemble in which fabric is specified.  The conservation of $W_{x}$, $W_{y}$ and $\mu_{klmn}$ are enforced via Lagrange multipliers $\lambda_{x}$, $\lambda_{y}$ and $\gamma_{klmn}$ respectively.  
\begin{eqnarray}
\frac{\partial}{\partial \nu_{p \ldots u}}\left[ \ln\Omega\{\nu_{i\ldots n}\}-\lambda_{x}\left(\sum_{i}\cdots\sum_{n}\nu_{i\ldots n}w_{xi}\right)\right. & & \nonumber\\*
-\lambda_{y}\left(\sum_{i}\cdots\sum_{n}\nu_{i \ldots n}w_{yj}\right) & & \nonumber\\*
-\left.\sum_{k}\cdots \sum_{n}\gamma_{k\ldots n}\left(\sum_{i}\sum_{j}\nu_{i\ldots n}\right)\right] & = & 0 \nonumber\\*
\ \label{max1}
\end{eqnarray}

The calculus is performed using Stirling's approximation and an expansion of the logarithm in a Taylor series of $\nu_{i \ldots n}$ where necessary.  Taking the limit for $S\to\infty$ while conserving $N$ and then taking the continuum limit obtains
\begin{eqnarray}
P_{2w4\theta}(w_{x},w_{y},\theta_{\beta}) & = & \Upsilon\left(w_{x},w_{y},\theta_{\beta}\right)\Psi\left(w_{x},w_{y},\theta_{\beta}\right)\nonumber\\*
& & \ \ \times G(\theta_{\beta})\ e^{-\lambda_{x}w_{x}-\lambda_{y}w_{y}}
\label{eqnp2w4th}
\end{eqnarray}
where the Fabric Partition Factor $G(\theta_{\beta})=G(\theta_{1},\ldots,\theta_{4})$ derives from $\exp(-\gamma_{p\ldots u})$ in the continuum limit.

Note that $\lambda_{x}$ and $\lambda_{y}$ should not be confused with the decay constants of the exponential tails in the empirical PDFs.  Most (if not all) of the exponential behavior in Eq.~(\ref{eqnp2w4th}) is contained in the form of $\Upsilon$ because it is a functional of $P_{F\theta}(F,\theta)$.  It will be shown in a numerical solution of the isotropic case using an approximation method (later in this paper) that the value of $\lambda_{x}=\lambda_{y}$ is approximately zero.  This should not be the case in general, however because these two parameters provide the only information about stress anisotropy in the equation.

The Fabric Partition Factor $G$, along with $\Upsilon$ and $\Psi$, determines the partition of fabric between the $(w_{x},w_{y})$ ``modes''.  Integrating Eq.~(\ref{eqnp2w4th}) over $w_{x}$ and $w_{y}$
\begin{eqnarray}
P_{4\theta}(\theta_{\beta}) & = & G(\theta_{\beta}) \int\!\!\!\int_{0}^{\infty}\!\!\!\text{d}^{2}w\ \Upsilon \Psi e^{-\lambda_{x}w_{x}-\lambda_{y}w_{y}} \nonumber\\*
& = & G(\theta_{\beta})\ H(\theta_{\beta}). \label{defineF}
\end{eqnarray}
Assuming the standard result of statistical mechanics, one form of $P_{F\theta}$ will be found in the overwhelming majority of the occupied phase space, and so in the thermodynamic limit we may treat fabric as if it is partitioned by $G$ with a fixed form in all of phase space.  This factor is not a function of $w_{x}$ or $w_{y}$.  However, the partition is not an equipartition because of the influence of $\Upsilon$ and $\Psi$.  The former is variable over the range of angle configurations within each mode, and the second is a truncating factor which limits the range of angle configurations differently for each mode.

\subsection{The Recursive ``Transport'' Equation}

The JPDF $P_{2w4\theta}$ can be collapsed,
\begin{eqnarray}
P_{F\theta}(F,\theta) & = & \frac{1}{4}\sum_{\beta=1}^{4}\int\!\!\!\int_{0}^{\infty}\!\!\!\text{d}^{2}w\!\!\!\int\!\!\!\int\!\!\!\int\!\!\!\int_{0}^{2\pi}\!\!\!\text{d}^{4}\theta\ \delta(\theta-\theta_{\beta})\nonumber\\*
& & \ \times \delta\left[F-F_{\beta}(w_{x},w_{y},\theta_{1},\ldots,\theta_{4})\right]\nonumber\\*
& & \ \times P_{2w4\theta}(w_{x},w_{y},\theta_{1},\ldots,\theta_{4})\label{collapse2}.
\end{eqnarray}
Inserting Eq.~(\ref{eqnp2w4th}) into Eq.~(\ref{collapse2}) results in a recursion of $P_{F\theta}$,
\begin{widetext}
\begin{eqnarray}
P_{F\theta}(F,\theta) = \frac{1}{4}\sum_{\beta=1}^{4}\int\!\!\!\int_{0}^{\infty}\!\!\!\text{d}^{2}w\ e^{-\lambda_{x}w_{x}-\lambda_{y}w_{y}}\ \int\!\!\!\int\!\!\!\int\!\!\!\int_{0}^{2\pi}\!\!\!\text{d}^{4}\theta\ \delta(\theta-\theta_{\beta})\ \delta\left[F-F_{\beta}(w_{x},w_{y},\theta_{1},\ldots,\theta_{4})\right]\nonumber\\*
\times\ G(\theta_{1},\ldots,\theta_{4})\ \prod_{\gamma=1}^{4} \Theta\left[F_{\gamma}(w_{x},w_{y},\theta_{1},\ldots,\theta_{4})\right]\ \int\!\!\!\int\!\!\!\int\!\!\!\int_{0}^{\infty}\!\!\!\!\!\text{d}^{4}F'\ \prod_{\delta=1}^{4} \left[P_{F\theta}\left(F'_{\delta},\theta_{\delta}\right)\right]^{\frac{1}{2}} \\*
\times\ \delta\left[w_{x}-w_{x}^{\text{right}}(F'_{1},\theta_{1},\ldots,F'_{4},\theta_{4})\right]\ \delta\left[w_{x}-w_{x}^{\text{left}}(\ \cdot\ )\right]\ \delta\left[w_{y}-w_{y}^{\text{top}}(\ \cdot\ )\right]\ \delta\left[w_{y}-w_{y}^{\text{bott.}}(\ \cdot\ )\right]\nonumber\label{transport}
\end{eqnarray}
\end{widetext}
This can be simplified by taking advantage of symmetries in the ensemble.  For example, in the case in which fabric is not quartered so that every contact $(\beta=1,\ldots,4)$ is statistically similar, then the summation may be removed.

The dependency of $\Upsilon$ upon the form of $P_{F\theta}(F,\theta)=P_{F\theta}(F,\theta \mid \{w_{\xi,\alpha},\theta_{\alpha \beta}\})$ reveals that the DOS in a granular ensemble is self-organized (accomplished by the packing in its dynamic state as it sought a stable locus in phase space) and cannot be given a simplistic \textit{a priori} characterization in a way that is analogous to the \textit{a priori} uniform characterization of a thermal DOS.  The form of $P_{F\theta}(F,\theta)$ derives from the DOS non-uniformity and \textit{vice versa}.  In principle, this recursion is the unique solution for this special case, assuming that correlative information is non-recursive in the ensemble average and that Edwards' flat measure over all metastable states is correct.

\subsection{The Mean Structure Approximation}

The recursion equation can be solved using Monte Carlo integration.  Efforts are underway to obtain the numerical solution, which shall be presented in a future publication.  For the present an approximation will be introduced, simplifying the recursion equation while yet providing sufficient accuracy to demonstrate the principle organizational features of the ensemble.  The approximation has value in its own right because it will isolate and identify those organizational features.

The approximation is obtained by projecting the DOS in $\mathbb{S}_{4} \to \mathbb{S}_{5}$, where the latter is the subspace $\{w_{x \alpha},w_{y \alpha}\}$. This projection is performed by integrating the DOS across all the angle axes.  For a given pair of values $(w_{x1},w_{y1})$, the evaluation of $\Upsilon(w_{x1},w_{y1},\theta_{\alpha \beta})\ne\Upsilon(w_{x1},w_{y1},\theta_{\gamma \delta})$ for $\{\theta_{\beta}\}_{\alpha} \ne \{\theta_{\delta}\}_{\gamma}$ in general.  That is, certain contact angle configurations $\left\{\theta_{\beta}\right\}$ will yield a greater probability that the grain will be consistent with their neighbors than will other contact angle configurations.  Therefore, information is lost in the projection into the $\mathbb{S}_{5}$ subspace.  Nevertheless, this loss of information may not be so great that it blurs the principle organization of the DOS.  Arguments can be advanced to show that, over the distribution of all contact angle configurations where the grain is stable,
\begin{equation}
\Upsilon(w_{x1},w_{y1},\theta_{\alpha \beta}) \approx \overline{\Upsilon}(w_{x1},w_{y1})\label{MSAstatement}
\end{equation}
for \textit{most} stable grain configurations.  Whereas $\Psi$ is a truncating factor in the DOS, defining the region where individual grains are stable, $\Upsilon$ is a scaling factor in the DOS, indicating how often particular grain configurations will occur in the ensemble based upon the probability that they can satisfy N3L with their neighbors.  Eq.~(\ref{MSAstatement}) claims that this scaling is strongly dependent upon the values of $w_{x}$ and $w_{y}$; but when varying the contact angles it does not vary too much over the majority of that configuration space.  This allows the $\mathbb{S}_{4} \to \mathbb{S}_{5}$ projection to be simplified.

\begin{figure}
\includegraphics[angle=0,width=0.45\textwidth]{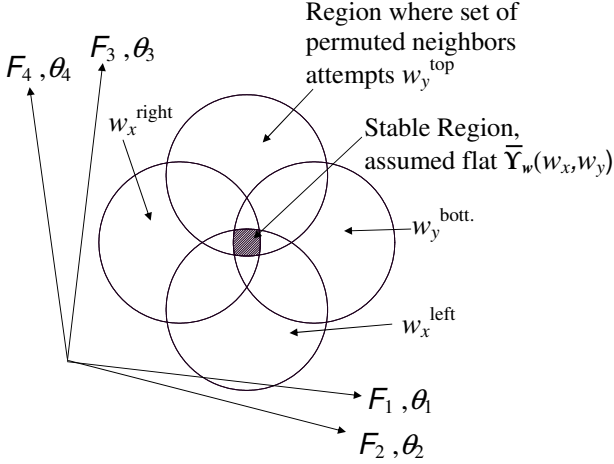}
\caption{\label{MSAdiagram} Schematic diagram of 8-dimensional space to illustrate the Mean Structure Approximation (MSA).  The MSA assumes that the probability for a grain to satisfy Newton's third law with its neighbors does not vary over any of the configurations of the grain having fixed Cartesian loads.  In fact, the exact probability does not vary too widely for the vast majority of those grain configurations.  Therefore, the MSA should produce a distribution of grain configurations that is a good representation of the exact ensemble.  The individual circles represent the region where random grain configurations, taken to be neighbors for the grain in question, would attempt to apply a particular load on each hemisphere of that grain.  The intersection is the stable region where the MSA applies.}
\end{figure}

The approximation shall be called the ``Mean Structure Approximation,'' or MSA, and it is illustrated in Fig.~\ref{MSAdiagram}.
It is important to distinguish this from the Mean Field Approximation (or MFA), which is useful for thermal systems but \text{not} acceptable for granular packings.  The reason that the MSA may be adequate where the MFA fails is because it preserves the exact intra-grain correlation of contact forces by N2L and also the approximate inter-grain correlations of $(w_{x \alpha},w_{y \alpha})$ by N3L, both of which are lost in the MFA.  The validity of the MSA shall be evaluated in the discussion section.

The most probable $P_{2w}(w_{x},w_{y})$ to occur in the $\mathbb{S}_{5}$ subspace can be obtained directly by integrating Eq.~(\ref{eqnp2w4th}) over all angles,
\begin{eqnarray}
P_{2w}(w_{x},w_{y}) & = & e^{-\lambda_{x}w_{x}-\lambda_{y}w_{y}}\ \int\!\!\!\int\!\!\!\int\!\!\!\int_{0}^{2 \pi}\!\!\!\!\text{d}^{4}\theta\ G(\theta_{\beta})\ \nonumber\\*
& & \times \Psi\left(w_{x},w_{y},\theta_{\beta}\right)\ \Upsilon\left(w_{x},w_{y},\theta_{\beta}\right) \nonumber\\*
\ \label{eqnp2w}
\end{eqnarray}
We wish to simplify this in the MSA.  With some manipulation it can be shown that,
\begin{eqnarray}
\Upsilon(w_{x},w_{y},\theta_{\beta}) & = & \int\!\!\!\int\!\!\!\int\!\!\!\int_{0}^{\infty}\!\!\!\!\!\text{d}^{4}w\left[P_{4w4\theta}^{\star}(w_{\xi}^{\text{hem.}},\theta_{\beta})\right]^{\frac{1}{2}}\nonumber\\*
& & \times\ \delta\left(w_{x}-w_{x}^{\text{right}}\right)\ \delta\left(w_{y}-w_{y}^{\text{top}}\right)\nonumber\\*
& & \times\ \delta\left(w_{x}-w_{x}^{\text{left}}\right)\ \delta\left(w_{y}-w_{y}^{\text{bott.}}\right) \nonumber\\*
\ \label{UpsdefY}
\end{eqnarray}
where,
\begin{eqnarray}
P_{4w4\theta}^{\star}(w_{\xi}^{\text{hem.}},\theta_{\beta}) & = & P_{4w4\theta}^{\star}(w_{x}^{\text{right}},\ldots,w_{y}^{\text{bott.}},\theta_{\beta}) \nonumber\\*
\ \nonumber\\*
& = & \int\!\!\!\int\!\!\!\int\!\!\!\int_{0}^{\infty}\!\!\!\!\!\text{d}^{4}F \ \prod_{\beta=1}^{4} P_{F\theta}\left(F_{\beta},\theta_{\beta}\right) \nonumber\\*
& & \times\ \delta\left(w_{x}^{\text{right}}-F_{1}\cos\theta_{1}-F_{4}\cos\theta_{4}\right)\nonumber\\*
& & \times\ \delta\left(w_{x}^{\text{left}}+F_{2}\cos\theta_{2}+F_{3}\cos\theta_{3}\right)\nonumber\\*
& & \times\ \delta\left(w_{y}^{\text{top}}-F_{1}\sin\theta_{1}-F_{2}\sin\theta_{2}\right)\nonumber\\*
& & \times\ \delta\left(w_{y}^{\text{bott.}}+F_{3}\sin\theta_{3}+F_{4}\sin\theta_{4}\right).\nonumber\\*
\ \label{Pstar4w4th}
\end{eqnarray}
This can be interpreted as the JPDF of \textit{attempted} loads and contact angles that the set of all possible packing permutations (with the specified $P_{F\theta}$) \textit{attempts} to place on any one grain.  The star indicates that this is only a conceptual distribution, not found in stable packings.  It can be viewed as a mean-field calculation, where the incoming forces have been drawn randomly from the entire set of grains in the packing permutations.  Its domain is therefore not restricted to the set of forces that would make a grain stable.  Because of this, the pair $(w_{x}^{\text{right}},w_{y}^{\text{top}})$ should not be too strongly correlated to $(w_{x}^{\text{left}},w_{y}^{\text{bott.}})$ after integrating out the angular dependence,
\begin{equation}\begin{array}{l}
P_{4w}^{\star}(w_{x}^{\text{right}},w_{y}^{\text{top}},w_{x}^{\text{left}},w_{y}^{\text{bott.}}) \approx \\*
\ \\*
\ \ \ \ \ \ \ \ \ \ \ \ \ \ \ \ P_{2w}^{\star}(w_{x}^{\text{right}},w_{y}^{\text{top}})P_{2w}^{\star}(w_{x}^{\text{left}},w_{y}^{\text{bott.}}). \end{array}\label{approx1}
\end{equation}

All the angular content of $\Upsilon$ is in $P_{4w4\theta}^{\star}(w_{\xi}^{\text{hem.}},\theta_{\beta})$, so we make the mean structure approximation,
\begin{equation}
P_{4w4\theta}^{\star}(w_{\xi}^{\text{hem.}},\theta_{\beta}) \approx P_{4w}^{\star}(w_{\xi}^{\text{hem.}})/16\pi^{4}
\end{equation}
for \textit{most} stable grain configurations, so that by Eq.~(\ref{UpsdefY}),
\begin{equation}
\Upsilon(w_{x},w_{y},\theta_{j}) \approx \left[P_{4w}^{\star}(w_{x},w_{x},w_{y},w_{y})\right]^{\frac{1}{2}}/4 \pi^{2}\label{approx2}
\end{equation}
for most $(w_{x},w_{y},\theta_{j})$.  Using Eqs.~(\ref{MSAstatement}), (\ref{approx1}), and~(\ref{approx2}) we define
\begin{equation}
\overline{\Upsilon}(w_{x},w_{y}) = P_{2w}^{\star}(w_{x},w_{y})/4 \pi^2.
\end{equation}
$P_{2w}^{\star}(w_{x},w_{y})$ may be viewed as a mean-field calculation of \textit{attempted} loads (for packing permutations having the specified $P_{F\theta}$), which the half-space of the ensemble attempts to place on the corresponding hemisphere of a grain (for each of the two Cartesian loads), and where the mean field includes the unstable regions of the phase space.  However, it must be emphasized that this is not a mean-field calculation of loads actually placed on the grains, but rather it is the approximate scale measuring how often particular modes will satisfy N3L and therefore occur in the ensemble.  The validity of the approximation depends on the relative weakness of $\Upsilon$'s dependence on the contact angles for most stable grain configurations.

Using the MSA in Eq.~(\ref{eqnp2w}),
\begin{eqnarray}
P_{2w}(w_{x},w_{y}) & = & e^{-\lambda_{x}w_{x}-\lambda_{y}w_{y}}\ \overline{\Upsilon}\left(w_{x},w_{y}\right) \nonumber\\*
& & \times \int\!\!\!\int\!\!\!\int\!\!\!\int_{0}^{2 \pi}\!\!\!\!\text{d}^{4}\theta\ G(\theta_{\beta})\ \Psi\left(w_{x},w_{y},\theta_{\beta}\right). \nonumber\\*
\ \label{eqnp2wapprox}
\end{eqnarray}
Defining,
\begin{eqnarray}
\overline{\Psi}(w_{x},w_{y}) & = & \int\!\!\!\!\int\!\!\!\!\int\!\!\!\!\int_{0}^{2 \pi}\!\!\!\!\text{d}^{4}\theta\ G(\theta_{\beta})\nonumber\\*
& & \ \times \prod_{\gamma=1}^{4}\Theta\Big[F_{\gamma}(w_{x},w_{y},\theta_{1},\dots,\theta_{4})\Big].\label{findpsi}
\end{eqnarray}
we may write,
\begin{equation}
P_{2w}(w_{x},w_{y}) = e^{-\lambda_{x}w_{x}-\lambda_{y}w_{y}}\ \overline{\Upsilon}\left(w_{x},w_{y}\right)\ \overline{\Psi}\left(w_{x},w_{y}\right)\label{eqnp2wapprox1}
\end{equation}

Finally, writing the DOS in $\mathbb{S}_{5}$, 
\begin{eqnarray}
\tilde{\rho}^{(5)}\{w_{x \alpha},w_{y \alpha}\} = N!\ \prod_{\alpha}\ \overline{\Upsilon}(w_{x \alpha},w_{y \alpha})\ \overline{\Psi}(w_{x \alpha},w_{y \alpha})\nonumber\\*
\ \ \ \times \delta\left(\sum_{\alpha} w_{x \alpha}- W_{x} \right)\ \delta\left(\sum_{\alpha} w_{y \alpha}- W_{y} \right).\ \label{DOS5}
\end{eqnarray}

We may identify $\overline{\Psi}(w_{x},w_{y})$ as the ``Grain Factor'' and $\overline{\Upsilon}(w_{x},w_{y})$ as the ``Structure Factor.'' These are the primary features of non-uniformity in the DOS.  Whereas $\overline{\Psi}$ derives from the configuration space of \textit{individual} grains (cohesionless N2L), $\overline{\Upsilon}$ derives from the configuration space of grains connecting together to form a packing structure (N3L).  These two factors were so-named because their separability (in the MSA) and their roles may be considered somewhat analogous to the separability and roles of the atomic form factor and structure factor of x-ray crystallography.

The meaning of $\overline{\Psi}$ can be illustrated easily through a change of variables.  We notice that for rigid, cohesionless grains there is no inherent force scale and hence stability must be independent of the overall scale of the forces.  Thus it is convenient to change variables,
\begin{equation}
s_{\alpha}=\frac{w_{x \alpha}-w_{y \alpha}}{w_{x \alpha}+w_{y \alpha}},\ \ \ \  t_{\alpha}=w_{x \alpha}+w_{y \alpha}.\label{stmap}
\end{equation}
The stability of the $\alpha^{\text{th}}$ grain is therefore a function of $s_{\alpha}$ and the four contact angles, $\{\theta_{\beta}\}_{\alpha}$, only.  With the Jacobian $J=t$, Eq.~(\ref{eqnp2w}) can also be written,
\begin{equation}
P_{st}(s,t)= \overline{\Upsilon}_{st}(s,t) \overline{\Psi}_{s}(s) e^{-(\lambda_{x}+\lambda_{y})t/2 -(\lambda_{x}-\lambda_{y})st/2 }\label{Pst}
\end{equation}
where the notation has been introduced,
\begin{eqnarray}
\overline{\Upsilon}_{st}(s,t) & = & t\ \overline{\Upsilon}\big[(1+s)t/2,(1-s)t/2\big]\nonumber\\*
& = & (w_{x}+w_{y})\ \overline{\Upsilon}(w_{x},w_{y}),
\end{eqnarray}
and,
\begin{eqnarray}
\overline{\Psi}_{s}(s) & = & \overline{\Psi}\big[(1+s)t/2,(1-s)t/2\big]\nonumber\\*
& = & \overline{\Psi}(w_{x},w_{y}).
\end{eqnarray}
Note that $\Theta$ in Eq.~(\ref{findpsi}) is insensitive to the scale of $F_{k}$ and cares only whether it is positive or negative, and hence the $t$ does not appear as an argument of $\overline{\Psi}_{s}(s)$.  

\begin{figure}
\includegraphics[angle=-90,width=0.45\textwidth]{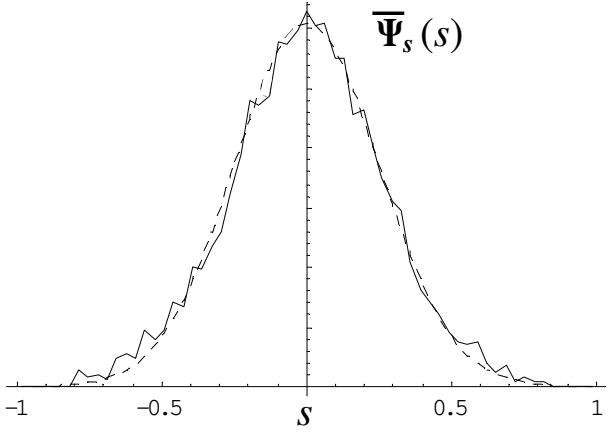}
\caption{\label{GrainFactor} Grain factor fit to Eq.~(\ref{PsiIsotropy}).}
\end{figure}

In these coordinates Eq.~(\ref{findpsi}) may be solved very efficiently by Monte Carlo integration.  This has been performed for the case of quartered isotropy.  In the MSA, $\overline{\Upsilon}$ does not affect the fabric partition, and hence it is easy to find the fabric partition factor.  The product of the weighting for the quartering bias with the weighting for the fabric partition was obtained empirically by adjusting as required in a Fourier decomposition to achieve approximate isotropy $P_{\theta}(\theta)\approx 1/2\pi$.  The numerical result for that case is fit well by a Gaussian,
\begin{equation}
\begin{array}{lr}\overline{\Psi}_{s}(s)=\sqrt{c/\pi}\ e^{-c s^2}, & |s|\le 1
\end{array}\label{PsiIsotropy}
\end{equation}
with $c=7.9$.  It is shown in Fig.~(\ref{GrainFactor}) with the fit as the dashed curve.  
This indicates that in the isotropic case the volume of a grain's stability space is a Gaussian function of the individual grain's load-anisotropy, $s$.

In contrast to the simplicity of the above result, the form of $\overline{\Upsilon}$ depends upon $P_{F\theta}$ and hence can only be found by solving the transport equation.

\subsection{The Mean Structure Transport Equation}

Just as Eq.~(\ref{eqnp2w4th}) can be solved recursively, giving us the recursive ``transport'' equation, so can Eq.~(\ref{eqnp2wapprox1}) be solved recursively, giving us the ``Mean Structure Transport Equation'' or MSTE.  

\begin{figure}
\includegraphics[angle=-90,width=0.45\textwidth]{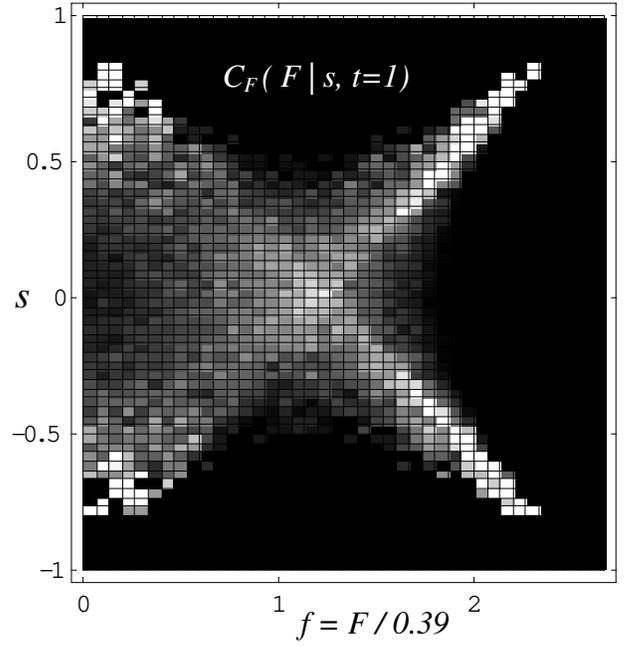}
\caption{\label{FMCondS1} Horizontal cross sections through this plot are the conditional PDF for $f$, the normalized contact force magnitudes ($\left<F\right>=0.39$ when $t=1$).  White represents higher probability density.  The vertical axis represents its dependence upon the $s$ variable with a fixed $t=1$.  Varying $t$ only rescales $f$.}
\end{figure}

To develop the MSTE, we convert the load distribution of Eq.~(\ref{eqnp2wapprox1}) into a contact force distribution.  This cannot be done simply by collapsing $P_{2w}$ since it does not contain sufficient information.  However, the variables may be changed if we first obtain the joint conditional PDF $C_{F\theta}\left(F,\theta \mid w_{x},w_{y}\right)$, so that,
\begin{equation}
P_{F\theta}(F,\theta)=\int\!\!\!\!\int_{0}^{\infty} \text{d}^{2}w \ \ C_{F\theta}(F,\theta \mid w_{x},w_{y})\ P_{2w}(w_{x},w_{y}).\label{transport2}
\end{equation}
This PDF can be obtained easily through the same change of variables introduced previously, $(w_{x},w_{y}) \to (s,t)$,  because $C_{F\theta}\left(F,\theta \mid s,t\right)= t \cdot C_{F\theta}\left(t F, \theta \mid s,1\right)$ and the conditional dependency is reducible to the $s$ variable, alone.  This may obtained by straightforward integration,
\begin{eqnarray}
C_{F\theta}\left(F,\theta\mid s,1\right) & = & \frac{1}{4}\sum_{\gamma=1}^{4}\int\!\!\!\!\int\!\!\!\!\int\!\!\!\!\int_{0}^{2 \pi}\!\text{d}^{4}\theta\ G(\theta_{\beta})\ \delta(\theta-\theta_{\gamma})\nonumber\\*
& & \ \ \times \delta\Big[F-F_{\gamma}(s,1,\theta_{1},\dots,\theta_{4})\Big]\nonumber\\*
& & \ \ \times \prod_{\eta=1}^{4}\Theta\Big[F_{\eta}(s,1,\theta_{1},\dots,\theta_{4})\Big]\ \label{findC}
\end{eqnarray}
where only one term of the sum is needed in many cases due to the symmetries of the ensemble.  This reflects the MSA because it assumes that all grains in the same $(s,t)$ mode contribute to the integral according to the same weight.  It can be found by very easy Monte Carlo integration, and the result for the case of isotropy, $C_{F\theta}(F,\theta \mid s,1) = C_{F}(F \mid s,1)/2\pi$, is shown in Fig.~(\ref{FMCondS1}).

Combining this PDF with Eq.~(\ref{eqnp2wapprox1}) and the definitions of $C_{F\theta}$, $\overline{\Psi}$ and $\overline{\Upsilon}$,
\begin{eqnarray}
P_{F\theta}(F,\theta) & = & \frac{1}{16 \pi^2}\sum_{\gamma=1}^{4}\int\!\!\!\!\int_{0}^{\infty} \text{d}^{2}w \int\!\!\!\!\int\!\!\!\!\int\!\!\!\!\int_{0}^{2 \pi}\!\text{d}^{4}\theta\ \ G(\theta_{\beta})\nonumber\\*
& & \times\ \delta(\theta-\theta_{\gamma})\ \delta\Big[F-F_{\gamma}(w_{x},w_{y},\theta_{1},\dots,\theta_{4})\Big]\nonumber\\*
& & \times \prod_{\eta=1}^{4}\Theta\Big[F_{\eta}(w_{x},w_{y},\theta_{1},\dots,\theta_{4})\Big] \nonumber\\*
& & \times\ e^{-\lambda_{x}w_{x}-\lambda_{y}w_{y}}\ P_{2w}^{\star}(w_{x},w_{y}). \nonumber\\*
\ \label{MSTEform}
\end{eqnarray}

$P_{2w}^{\star}(w_{x},w_{y})$ used in this equation may be obtained a number of ways that should be equivalent within the accuracy of the MSA.  Two of these have been used in the numerical results and were shown indeed to produce identical results to within the statistical precision of the data.  The first is purely consistent with the MSA, assuming no necessity for \textit{a priori} correlation between the loads and the contact angles.  Furthermore, it assumes no \textit{a priori} correlation between $w_{x}$ and $w_{y}$.  Correlations arise only after throwing out unstable grain configurations.  That is, it assumes a fixed $\overline{\Upsilon}$ over the union of two circles in Fig.~(\ref{MSAdiagram}), not just the intersection of all four (the gray region).  Imposing $\overline{\Psi}$ then throws out grain configurations outside of the gray region.  This method is,
\begin{eqnarray}
w_{x} & = & F_{1}\cos\theta_{1}+F_{2}\cos\theta_{2},\nonumber\\*
w_{y} & = & F_{3}\sin\theta_{3}+F_{4}\sin\theta_{4}
\end{eqnarray}
(note that all four contacts are treated as if distinctly different despite the fact that an $x$-hemisphere and a $y$-hemisphere overlap in one quadrant), and
\begin{equation}
P_{2w}^{\star}(w_{x},w_{y}) = P_{wx}^{\star}(w_{x}) P_{wy}^{\star}(w_{y}),
\end{equation}
where
\begin{eqnarray}
P_{wx}^{\star}(w_{x}) & = & \int\!\!\!\!\int_{0}^{\infty}\!\!\!\text{d}^{2}F\ \int\!\!\!\!\int_{0}^{2\pi}\!\!\!\text{d}^{2}\theta\ \prod_{\beta=1}^{2} P_{F\theta}\left(F_{\beta},\theta_{\beta}\right)\nonumber\\*
& & \ \ \ \times\ \delta(w_{x}-F_{1}\cos\theta_{1}-F_{2}\cos\theta_{2}),\nonumber\\
\ \\
P_{wy}^{\star}(w_{y}) & = & \int\!\!\!\!\int_{0}^{\infty}\!\!\!\text{d}^{2}F\ \int\!\!\!\!\int_{0}^{2\pi}\!\!\!\text{d}^{2}\theta\ \prod_{\gamma=3}^{4} P_{F\theta}\left(F_{\gamma},\theta_{\gamma}\right)\nonumber\\*
& & \ \ \ \times\ \delta(w_{y}-F_{3}\sin\theta_{3}-F_{4}\sin\theta_{4}).\nonumber\\*
\ \label{method1}
\end{eqnarray}

The second method, which will also be used in a Monte Carlo solution of the PDFs, attempts greater fidelity to the micromechanics by imposing \textit{a priori} correlation between $w_{x}$, $w_{y}$ and $\{\theta_{\beta}\}$.  If the MSA is valid, then imposing these correlations should be largely superfluous.  Comparing the results of these two methods will therefore test the MSA in Sec.~IV.A.  The second method, which for simplicity is expressed here for the case of quartered fabric, is
\begin{eqnarray}
w_{x} & = & F_{1}\cos\theta_{1}+F_{2}\cos\theta_{2},\nonumber\\*
w_{y} & = & F_{2}\sin\theta_{2}+F_{3}\sin\theta_{3}
\end{eqnarray}
(note the shared contact $\vec{F}_{2}$), and
\begin{eqnarray}
P_{2w3\theta}^{\star}(w_{x},w_{y},\theta_{\beta}) & = & \int\!\!\!\!\int\!\!\!\!\int_{0}^{\infty}\!\!\!\text{d}^{3}F\ \prod_{\gamma=1}^{3} P_{F\theta}\left(F_{\gamma},\theta_{\gamma}\right)\nonumber\\*
& & \times\ \delta(w_{x}-F_{1}\cos\theta_{1}-F_{2}\cos\theta_{2})\nonumber\\*
& & \times\ \delta(w_{y}-F_{2}\sin\theta_{2}-F_{3}\sin\theta_{3}).\nonumber\\*
\ \label{method2}
\end{eqnarray}

Inserting either of these forms of $P_{2w}^{\star}$ into Eq.~(\ref{MSTEform}) produces an MSA recursion equation in $P_{F\theta}$, which is the MSTE.  It can be simplified by taking advantage of the various symmetries of the ensemble.

The two different forms of $P_{2w}^{\star}$ produce two different forms of the MSTE.  This is striking because one form of $P_{2w}^{\star}$ contains $(P_{F\theta})^{3}$ whereas the other contains $(P_{F\theta})^{4}$.  The ability of these two very different transport equations to produce the same $P_{F\theta}$ depends upon the validity of the MSA.

\section{Results}

Here, the following nomenclature is used.  The vector magnitude of the contact forces are denoted by $F$, their distribution is $P_{F}(F)$, and their mean is $\left<F\right>$.  The corresponding normalized force magnitudes are $f=F/\left<F\right>$, which have a distribution $P_{f}(f)=\left<F\right>P_{F}(f\left<F\right>)$.  The Cartesian force components in the $x$ direction are denoted by $F_{x}$, their distribution is $P_{X}(F_{x})$, and their mean is $\left<F_{x}\right>$.  The corresponding normalized Cartesian forces are $f_{x}=F_{x}/\left<F_{x}\right>$, which have a distribution $P_{x}(f_{x})=\left<F_{x}\right>P_{X}(f_{x}\left<F_{x}\right>)$.  

The MSTE in the previous section was solved in a Monte Carlo process for the case of isotropic stress and fabric, with one further simplification.  It was found that $\lambda_{x}$ and $\lambda_{y}$ were not exactly zero in the MSA, although they were very tiny $\sim 0.01$ so that the exponential factors were not exactly unity but were nevertheless negligible.  Therefore, rather than implementing the exponential weighting exactly, the forces were simply rescaled with a flat factor in each iteration to prevent incremental growth.  This approach is reasonable because the phase space for rigid grains has no inherent force scale, the growth was very small, and the growth was balanced in the $x$ and $y$ components.  Hence, the form of the DOS should not be greatly affected by this flat rescaling.

The MSTE was shown to converge efficiently to the same stationary state after beginning from several different initial distributions.  The original work was performed with \textit{Mathematica}\textcircled{\tiny{R}} solving for approximately $5,500$ grains.  These results are presented in this paper.  Ongoing efforts with Fortran demonstrate that converged solutions can be found for a million contacts in about 1 minute on a desktop computer.  It is quite easy to obtain data sets of $10^{10}$ grains or greater, making it possible to study joint or conditional distributions of three or more variables with smooth statistics using only a desktop computer.  For some applications this provides a tremendous computational advantage over the fully dynamic simulations.

The $P_{f}\left(f\right)$ resulting from the transport method was shown earlier in Fig.~(\ref{fig:vectorPDF}).  It has all the key characteristics of granular contact force PDFs.  A fit, shown as the smooth curve in Fig.~(\ref{fig:vectorPDF}), was obtained with the form proposed for the data from the carbon-paper experimental method \cite{chicago},
\begin{equation}
P_{f}\left(f\right)= a \Big(1- b e^{-c f^{2}}\Big)e^{-d f}.
\label{eqn:chicagofit}
\end{equation}
Using the values $a=3.28$, $b=0.85$, $c=1.56$, and $d=1.56$, the fit is excellent and is in quantitative agreement with the range of values reported from both experiments and numerical simulations.  It should be noted that here, as in most of the empirical distributions \cite{chicago, silbertgrest, radjaijean}, $d$ is suspiciously close to $\pi/2$.  A plausible reason why this value arises under isotropic conditions is provided in the discussion section.

\begin{figure}
\includegraphics[angle=0,width=0.45\textwidth]{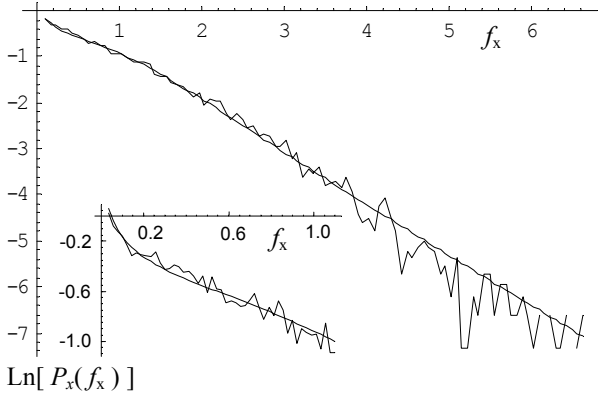}
\caption{\label{fig:CartesianPDF} Semi-logarithmic plot of the PDF $P_{x}\left(f_{x}\right)$ of the normalized $x$-components of the granular contact forces $f_{x}=F_{x}/\left<F_{x}\right>$.  The smooth curve was obtained from Eq.~(\ref{eqn:integral}).  The semi logarithmic inset shows the behavior below $f_{x}=1$.}   
\end{figure}

For the special case of true isotropy in which
\begin{eqnarray}
P_{F\theta}(F,\theta) & = & P_{F}(F) P_{\theta}(\theta)\nonumber\\*
& = & P_{F}(F)/2 \pi,
\end{eqnarray}
changing variables to Cartesian components $F_{x} = F \cos\theta$ is effected in probability theory by the straightforward
\begin{equation}
P_{X}\left(F_{x}\right) = \int_{0}^{2 \pi}\!\!\!\text{d}\theta\int_{0}^{\infty}\!\!\!\text{d}F\ \frac{P_{F}\left(F\right)}{2 \pi} \delta\left(F_{x}-F \cos\theta\right),
\end{equation}
or by evaluating the inner integral and expressing as normalized forces,
\begin{equation}
P_{x}\left(f_{x}\right) = \frac{2}{\pi}\frac{\left<F\right>}{\left<F_{x}\right>}\int_{0}^{\frac{\pi}{2}}\!\!\!\text{d}\theta\ P_{f}\left(f_{x} \sec\theta\right) \sec\theta,
\label{eqn:integral}
\end{equation}
where the symmetries of isotropy were used to reduce the range of integration in $\theta$.  Numerically integrating this \cite{snoeih2} with the $P_{f}\left(f\right)$ of Eq.~(\ref{eqn:chicagofit}) yields the smooth line in Fig.~(\ref{fig:CartesianPDF}).
It fits the numerical Cartesian component data from the transport algorithm (shown in the same figure) over the entire range.  It has a singularity at $f_{x}=0$ and is monotonically decreasing as demonstrated in numerical simulations \cite{snoeih1, snoeih2}.  It is not purely exponential, the two knees being indicative of a summation of $n^{\text{th}}$ order Modified Bessel Functions of the Second Kind, $K_{n}(\beta_{x} f_{x})$, functions which result naturally when exponential forms are used for $P_{f}\left(f\right)$ in Eq.~(\ref{eqn:integral}).  

The only problem with the fit shown in Fig.~(\ref{fig:vectorPDF}) occurs in the region of very small forces, $f\lesssim 0.2$.  This is the same region in which the form of Eq.~(\ref{eqn:chicagofit}) could not be experimentally verified due to calibration limits.  Therefore it is not known whether this is the correct empirical form in that region \cite{boundary}.  A better fit can be obtained using another form so that it fits excellently over the entire range including $f<<1$.  This will be accomplished starting with the observation noted above, that the two knees in Fig.~(\ref{fig:CartesianPDF}) are indicative of $K_{n}(\beta_{x} f_{x})$.  These two knees appear very dramatically in a rotation of the coordinates, a rotation which is most easily understood if performed manually by lifting the edge of the page toward the eye and rotating it so that the line of sight is parallel to the segments of the graph in Fig.~(\ref{fig:CartesianPDF}).  The fit to $P_{f}(f)$ will therefore be accomplished by fitting the natural forms to $P_{x}\left(f_{x}\right)$ and then mathematically inverting the transformation of Eq.~(\ref{eqn:integral}).  The simplest fit to within the statistical accuracy of this data set appears to be of the form,
\begin{equation}
P_{x}\left(f_{x}\right)= C_{1} \sum_{n=0}^{2} a_{n} f_{x}^{n} K_{n}\left(\beta_{x}f_{x}\right)
\label{eqn:besselfit}
\end{equation}
with $a_{0}=2$, $a_{1}=-2$, $a_{2}=11$, and $\beta_{x}=\pi/2$, and where $C_{1}$ is for normalization.
\begin{figure}
\includegraphics[angle=0,width=.45\textwidth]{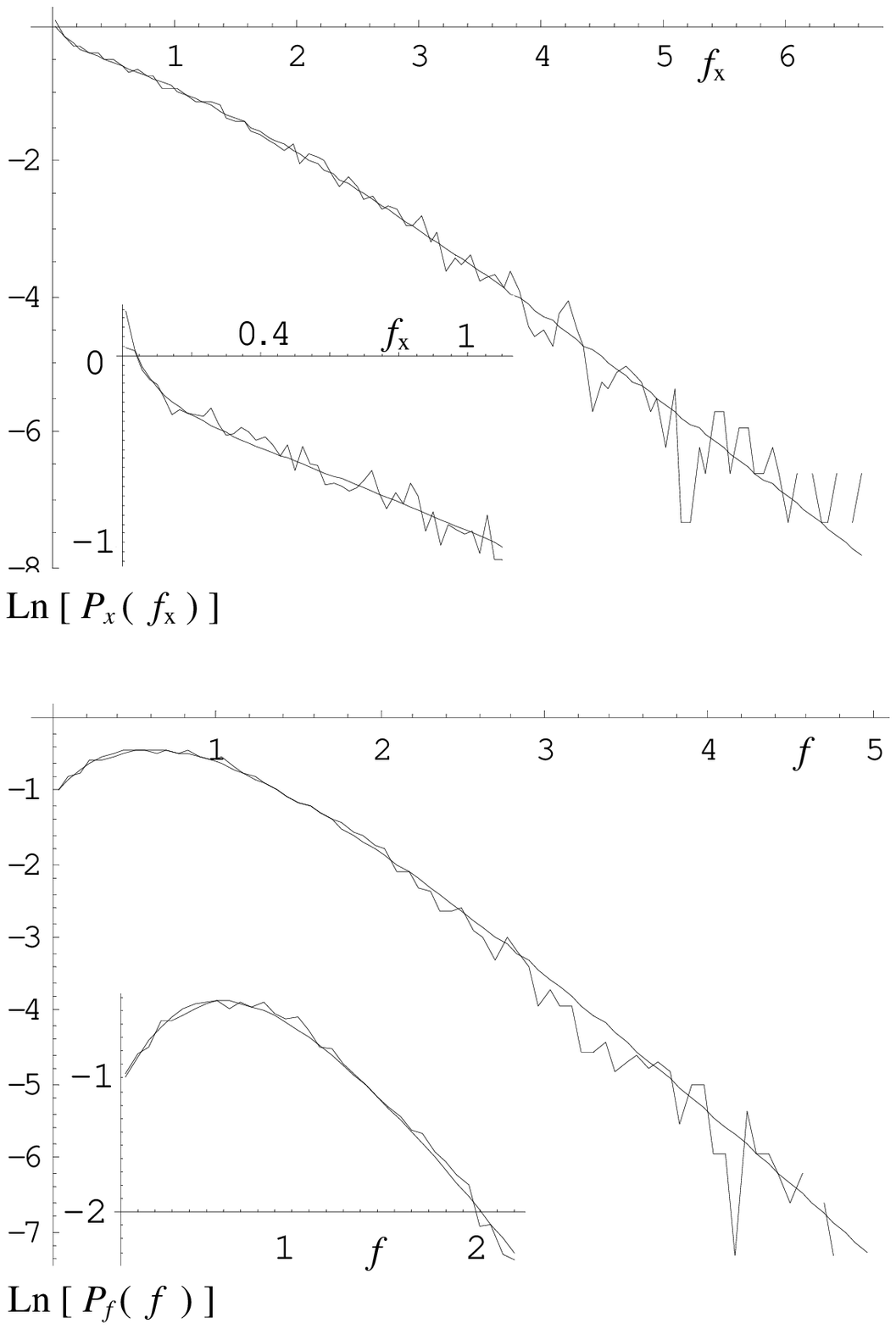}
\caption{\label{LogPair} (Top) The normalized Cartesian force components $f_{x}$ from the Mean Structure Transport Method fitted to Eq.~(\ref{eqn:besselfit}), which appears to be the natural form.  The inset shows the behavior below $f_{x}=1$.  (Bottom)  The force magnitudes $f$ from the Mean Structure Transport Method fitted to Eq.~(\ref{eqn:fbesselfit}).  The inset shows the behavior below $f=2$.  These two fits analytically transform to one another through Eq.~(\ref{eqn:integral}) and Eq.~(\ref{eqn:inverse}).}
\end{figure}
The fit is excellent over the entire range, displaying all the correct knees and piecewise slopes as shown in Fig.~(\ref{LogPair})(top).  The shape of the knee closest to $f_{x}=0$ could be obtained only by including a $K_{0}$ term.  This term has infinite probability density for $f_{x}=0$, but the singularity is very narrow and hence cannot usually be seen in a finite set of empirical data that has been aggregated into bins of finite width \cite{converge}.  

The transformation integral which is the inverse of Eq.~(\ref{eqn:integral}) cannot be deduced by probability theory because $f_{x}$ and $\theta$ are not statistically independent.  Therefore, inverting the change of variables to go from $(f_{x},\theta) \to (f,\theta)$ is not trivial, even in this isotropic case.  Nevertheless, the exact relationship can be derived using an approach which is equivalent to the mathematics of X-Ray Tomography \cite{youngquist}.  The result is,
\begin{equation}
P_{F}\left(F\right)=\frac{1}{F}\int_{0}^{\frac{\pi}{2}}\!\!\!\text{d}\theta\ P_{X}\left(F \sec\theta\right) \csc^{2}\theta,
\end{equation}
or, in normalized forces,
\begin{equation}
P_{f}\left(f\right)=\frac{\left<F_{x}\right>}{\left<F\right>}\frac{1}{f}\int_{0}^{\frac{\pi}{2}}\!\!\!\text{d}\theta\ P_{x}\left(f \sec\theta\right) \csc^{2}\theta.\label{eqn:inverse}
\end{equation}
This relationship is fascinating because we know that $F_{x}=F \cos\theta$ and therefore $F_{x} \le F$ for all $\theta$; however, this relationship computes $F$ in terms of $F_{x}=F\sec\theta$ so that $F_{x} \ge F$ for all $\theta$.  This says that the probability of finding a contact force magnitude $F$ is a weighted sum over the probabilities for all the Cartesian components $F_{x}$ that are too large to be relevant.  Nevertheless it is mathematically correct.

\begin{figure}
\includegraphics[angle=0,width=.45\textwidth]{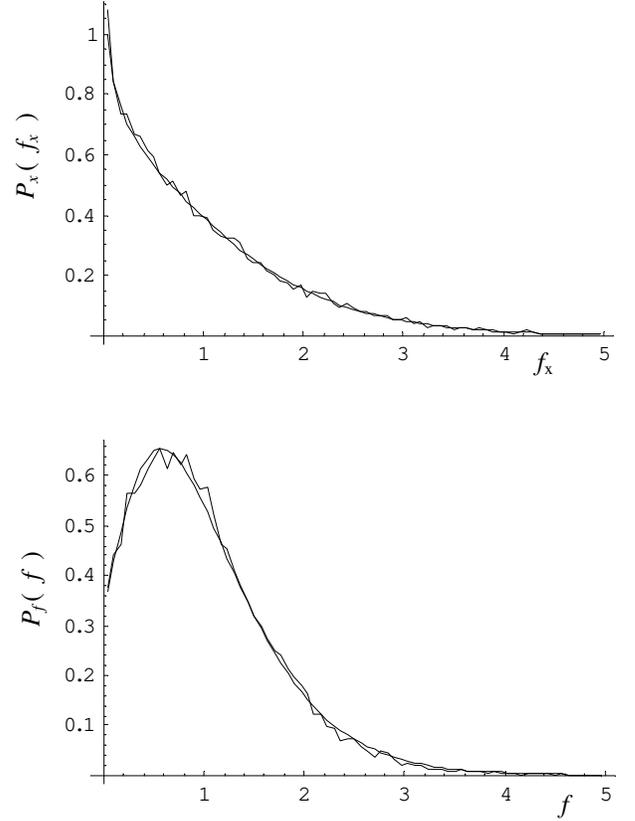}
\caption{\label{LinPair} (Top) Linear plot of the normalized Cartesian force components $f_{x}$ from the MSTE fitted to Eq.~(\ref{eqn:besselfit}).  (Bottom)  Linear plot of the force magnitudes $f$ from the MSTE fitted to Eq.~(\ref{eqn:fbesselfit}).}
\end{figure}

Using Eq.~(\ref{eqn:besselfit}) in Eq.~(\ref{eqn:inverse}), we obtain,
\begin{equation}
P_{f}\left(f\right)=\frac{\pi C_{2}}{2}e^{-\beta f}\sum_{n=0}^{2} b_{n} \left<F\right>^{n} f^{n}
\label{eqn:fbesselfit}
\end{equation}
with $C_{2} = C_{1}$, $b_{0} = a_{0}$, $b_{1} = \pi a_{1}/2+a_{2}$, $b_{2} = \pi a_{2}/2$, and $\beta = \beta_{x}\left<F\right>/\left<F_{x}\right>\approx (\pi/2)^{2}$.  This result fits the numerical data from the MSTE excellently over the entire range of $f$ as shown in Fig.~(\ref{LogPair})(bottom).  It exactly matches the finite and nonzero value of $P_{F}\left(0\right)= \frac{\pi}{2}C_{1} a_{0}$ that occurred in the numerical data, so we see that the $a_{0}$ term that made $P_{x}\left(f_{x} \to 0\right)$ infinite is the same $b_{0}$ term that makes $P_{f}\left(0\right)$ nonzero and finite.  The linear plots of Eqs.~(\ref{eqn:besselfit}) and~(\ref{eqn:fbesselfit}) are shown in Fig.~(\ref{LinPair}) in order to show that the curve fits are truly good in the region of weak forces, even without the compression of a logarithmic axis.

\begin{figure}
\includegraphics[angle=-90,width=.45\textwidth]{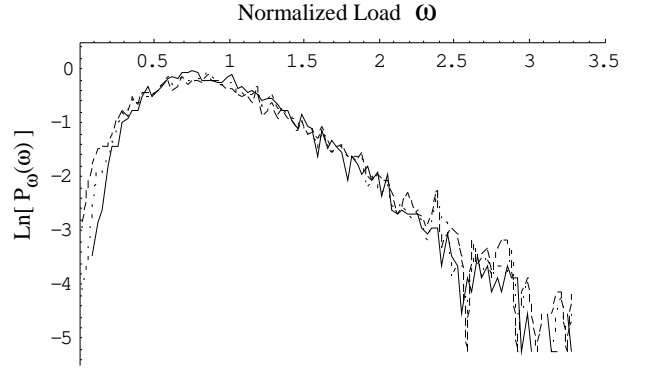}
\caption{\label{fig:weight} Semi logarithmic plot of the Cartesian Load PDF, the PDF of the total normalized load borne by each grain in the $x$ or $y$ direction unrotated (solid line), rotated $\pi/6$ radians (dotted line), and $\pi/4$ radians (dashed line).}
\end{figure}

Fig.~(\ref{fig:weight}) shows semi-logarithmically the Cartesian Load PDF $P_{w}(w)$ produced by the MSTE, computed for several different rotations of the Cartesian axes. 
These distributions have an exponential tail and a peak near $w \approx 1$.  The near similarity of the rotated plots indicates approximate rotational symmetry for this nearly isotropic model, despite its quartered fabric.  The variation in the region of weak loads is the result of that quartering.  In the unrotated axes, wherein the grains have exactly two contacts on each hemisphere, we find $P_{w}(w) \to 0$ as $w \to 0$.  We may fit $P_{w}(w)$ in these unrotated axes to an exponential with a power law prefactor,
\begin{equation}
P_{w}\left(w_{x}\right)=\left(\frac{w_{x}}{\left<w_{x}\right>}\right)^\alpha e^{-\beta w_{x}/\left<w_{x}\right>}.
\end{equation}
If the distribution of $F_{x}$ had been purely exponential and if there had been no correlation between adjacent values of $F_{x}$ on the same grain, then this should have had values of $\alpha=1.0$, $\beta=2.0$, and $\left<w_{x}\right>=2\left<F_{x}\right>$ as in the uniform $q$ model.  We do find an excellent fit over the entire curve using this form, and we do find that $\left<w_{x}\right>=2.0\ \left<F_{x}\right>$, but the fit is obtained with the values $\alpha=3$ and $\beta=4$.

By comparison, when the Cartesian axes are rotated the grains in this model may have 1, 2 or 3 contacts on the sampled hemisphere instead of the strict 2 contacts per hemisphere (1 contact per quadrant) that was defined for the unrotated axes.  The $P_{w}(w)$ for these rotated axes are also shown in Fig.~(\ref{fig:weight}).  They begin with a \textit{finite} probability density for zero force instead of beginning at zero, and the finite value is maximized when the rotation is $\pi/4$ radians because this is where we obtain the maximum fraction of grains having something other than 2 contacts on the hemisphere.  It was found in numerical simulations \cite{snoeih1, snoeih2} that when the grains in the bulk are segregated into separate populations having one, two, or three contacts on one side of the grain, respectively, then the Cartesian weight on the grains which support two or three others has a $P_{w}(w)$ which does go to zero probability for $w\to0$.  It is the population which supports only one contact which has a nonzero $P_{w}(w)$ because the load in that case is closely related to $P_{f}(f)$, which itself is nonzero at zero force.  Thus, the MSTE results are in agreement with this aspect of the simulation data, as well.

\begin{figure*}
\includegraphics[angle=-90,width=\textwidth]{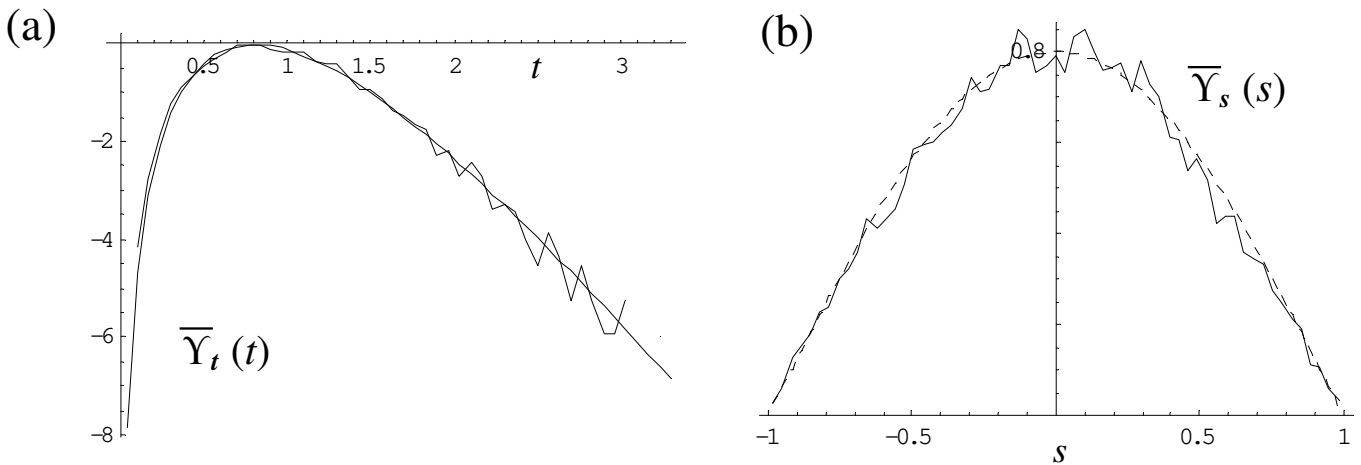}
\caption{\label{StructureFactor} Structure factor obtained from the Mean Structure Transport Method, fit to Eq.~(\ref{Upsilon}) for the case of isotropy with $\left<w_{x}\right>=\left<w_{y}\right> =\left<t\right>/2 =1/2$,  (a) semilog plot of $t$-dependence, (b) linear plot of $s$-dependence.}
\end{figure*}

The distribution of $s$ and $t$ variables resulting from the transport method are fit excellently by
\begin{equation}
P_{st}(s,t)=A \cos\left(\frac{\pi}{2}s\right)\left(\frac{t}{\left<t\right>}\right)^{4} e^{-5t/\left<t\right>}e^{-7.9s^2}.
\end{equation}
Thus, by comparing Eqs.~(\ref{PsiIsotropy}) and~(\ref{Pst}) with $\lambda_{x}=\lambda_{y}=0$, the structure factor can be identified,
\begin{eqnarray}
\overline{\Upsilon}_{st}(s,t) & = & \cos\left(\frac{\pi}{2}s\right)\left(\frac{t}{\left<t\right>}\right)^4 e^{-5t/\left<t\right>}\nonumber\\*
& = & \overline{\Upsilon}_{s}(s)\ \overline{\Upsilon}_{t}(t)
.\label{Upsilon}
\end{eqnarray}
$\overline{\Upsilon}_{t}$ and $\overline{\Upsilon}_{s}$ resulting from the transport method are shown in Fig.~(\ref{StructureFactor}) with smooth curves from Eq.~(\ref{Upsilon}).  

\section{Discussion}

\subsection{Validity of the Approximations}

The two approximations which enabled this ensemble analysis are the First Shell Approximation (FSA) and the Mean Structure Approximation (MSA).  Ultimately, the quantitative validation of these requires a careful comparison with numerical simulation data for particular states of the stress, fabric, and rheological history, and this has not yet been performed.  Meanwhile, the qualitative validity is already evident as discussed below.  

\subsubsection{Validity of the FSA}

Beside the constraints which defined the ensemble's DOS Eq.~(\ref{DOS2}), another geometric constraint is needed to ensure closure of every ``loop'' of grains in a packing.  Without this closure, the chains of contacting grains are allowed to branch out ever increasingly in all directions and overlap into one another's space.  Geometrically, then, omitting this constraint does not produce a good approximation to a packing.  However, it may still be an excellent approximation as far as the statistics of single-grain states are concerned.

It has been shown \cite{silbertgrest} that contact forces on the same grain are strongly correlated with one another.  There is anti-correlation for contacts closer together than $\Delta\theta \approx 0.4 \pi$ radians of angular separation, and a positive correlation when the angular separation is greater than that.  The correlation continues to increase as the contacts are increasingly distant from one another but still on the same grain.  The correlation dramatically drops immediately thereafter when the distance between contacts becomes greater than one grain diameter.  

\begin{figure}
\includegraphics[angle=0,width=0.30\textwidth]{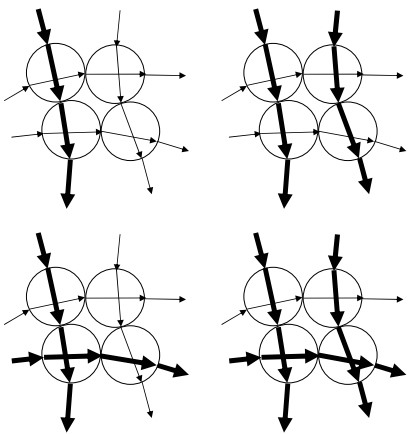}
\caption{\label{quadpacks} Contacts that are approximately $\pi/2$ radians away from one another on the same grain are only weakly correlated as illustrated by the closed loop of four grains that allows any combination of weak and strong force chains to pass through it.  If the angles were precisely $\pi/2$, then the four force chains in this figure would be completely independent.}
\end{figure}

The strong intra-grain relationships make sense due to the requirements of static equilibrium of the individual grains.  Contacts on the same quadrant compete for a share of the same load and hence are anti-correlated.  Contacts opposite one another transmit load through one another and hence are correlated.  Simplistically we could expect $\Delta\theta = \pi/2$ to be the crossover point of no correlation as illustrated in Fig.~(\ref{quadpacks}).
This is approximately correct, and the error is probably attributable to the existence of three-grain loops, history-dependent frictional effects, and so on.

Likewise, the sudden drop in correlation after one grain diameter of separation is also understandable in terms of the local mechanics.  It is true that neighboring grains share a common contact so that contacts on adjacent grains are just two sequential two-point correlations away from one another.  This induces correlations between them.  However, these inter-grain correlations should be primarily the result of the information contained in the sequential two-point intra-grain correlations because the lack of cohesion makes the grains otherwise (largely) independent.  Additional constraints are not found in the packing until entirely closed loops of grains are considered so that the sequential two-point correlations come all the way around the loop back to the original grain, again.  In 2D the typical closed loop consists of four grains, each grain being a vertex between a pair of contacts that form the loop.  The four-point correlation constructed as three sequential two-point correlations going the long way around a loop would undoubtedly be very weak compared to the single two-point correlation going the short way around the same loop, since the short way is intra-grain.  Hence, the extra correlation information imposed going the long-way around the loop must be very weak compared to the information already present intra-grain.  It should therefore be an excellent approximation to neglect this additional information and consider only the intra-grain relationships in defining the DOS.  This is the essence of the FSA.

This is not a rigorous argument because we should consider the sum of information from \textit{all} the loops in the packing that contain the grain in question, and it is conceivable that the sum of very many weak contributions may be strong.  However, due to the randomness of the packing, and the large number of amorphous packings that may exist in the configuration space, it is expected that the contributions from increasingly larger loops of grains will be increasingly decoherent and largely cancel one another.  Hence, there is good reason to assume that only the intra-grain contribution to the correlations is significant in agreement with the FSA.

If correct, the FSA is an important statement of the physics because it fundamentally characterizes the DOS and provides deep insight into the organization of the physics.  In contrast to thermal systems, with granular packings it would be completely unsatisfactory to use a mean-field approximation because this would throw away the structure resulting from the strong two-point correlations (remembering that these have been observed empirically).  However, by including only this next higher level in the approximation, that is, only the two-point correlations (and assuming that higher correlations exist strictly as a sequence of two-point correlations) the maximization of a state-counting entropy and the solution of the resulting transport equation produces excellent results as shown in the previous Sec.~III.  The two-point correlations therefore appear to be the essence of the physics.  Further work is needed to carefully test this hypothesis.

\subsubsection{Validity of the MSA}

The MSA is important because, if correct, it characterizes the structure factor as being a functional of $P_{2w}^{\star}(w_{x},w_{y})$ rather than $P_{F\theta}(F,\theta)$, and this offers the possibility to decouple the fabric from the force distributions in a way that will help the development of a full theory of rheology.  In the meantime, pending rigorous testing of the MSA, the following three considerations are presented to help justify it.

First, the results produced by the MSA appear to be in excellent agreement with the numerical simulation data.  A focused effort is needed to further test the quantitative agreement in specific cases of stress and fabric.

\begin{figure}
\includegraphics[angle=0,width=.45\textwidth]{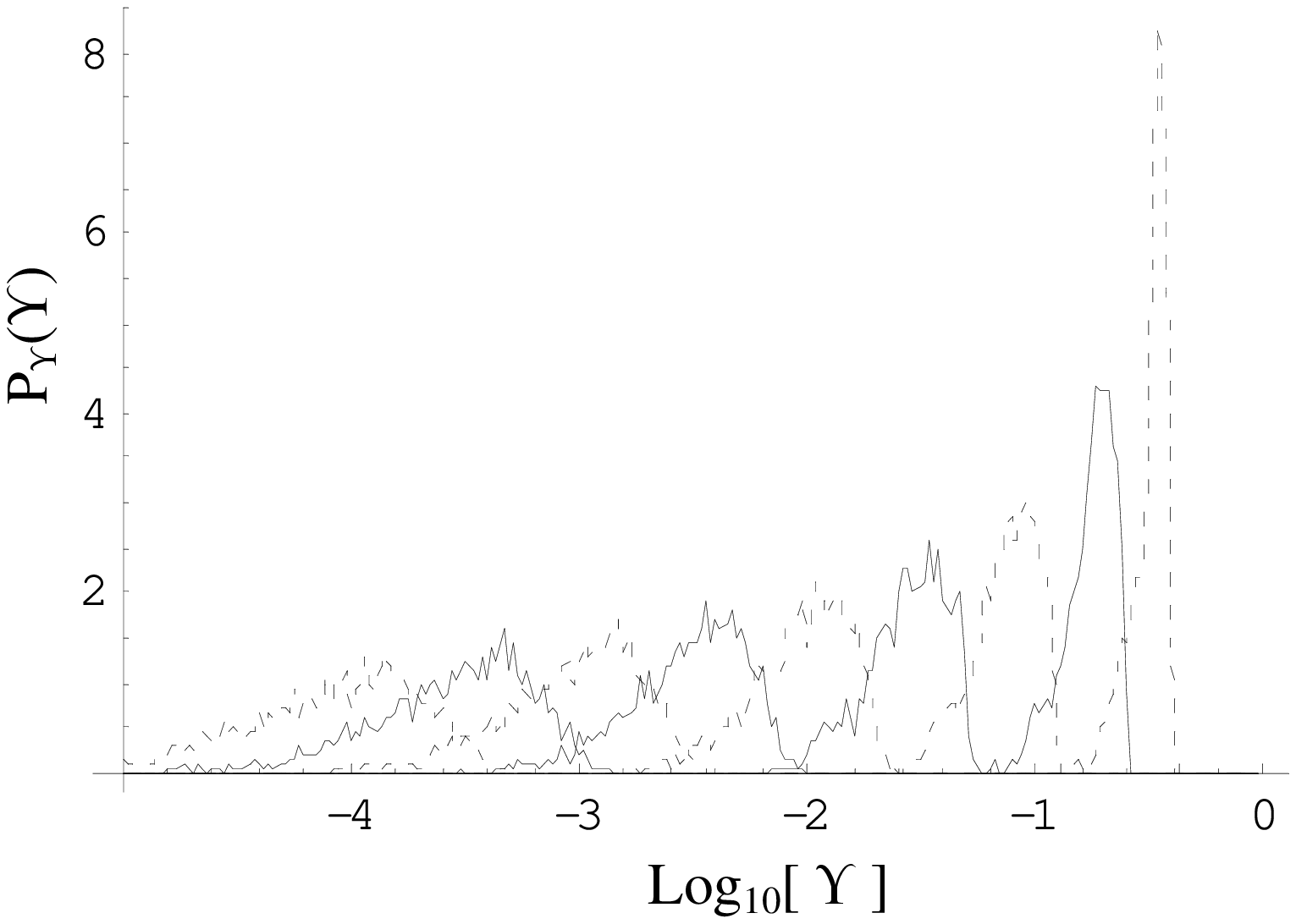}
\includegraphics[angle=0,width=.45\textwidth]{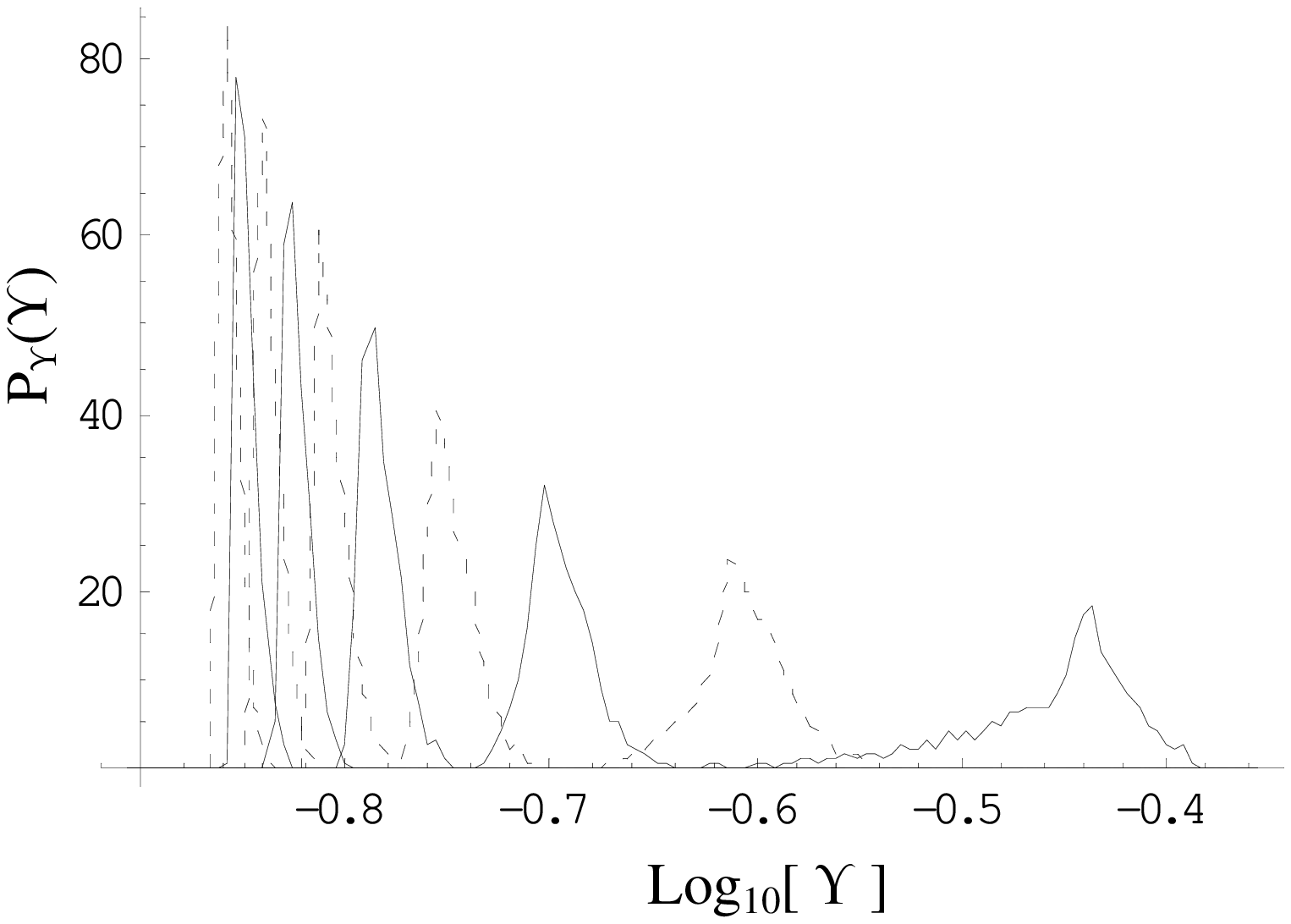}
\caption{\label{prob_ups_s0} Distribution of values of $\Upsilon(s,t,\theta_{j})$ for fixed value $s=0$ and several fixed values of $t$.  (Top) From left to right, $t=10$ (dashed), $9$ (solid), $8$ (dashed), $7$ (solid), $6$ (dashed), $5$ (solid), $4$ (dashed), $3$ (solid), and $2$ (dashed).  (Bottom) From left to right, $t=1/10$ (dashed), $1/9$ (solid), $1/8$ (dashed), $1/7$ (solid), $1/6$ (dashed), $1/5$ (solid), $1/4$ (dashed), $1/3$ (solid), $1/2$ (dashed), and $1$ (solid).}
\end{figure}

\begin{figure}
\includegraphics[angle=0,width=.45\textwidth]{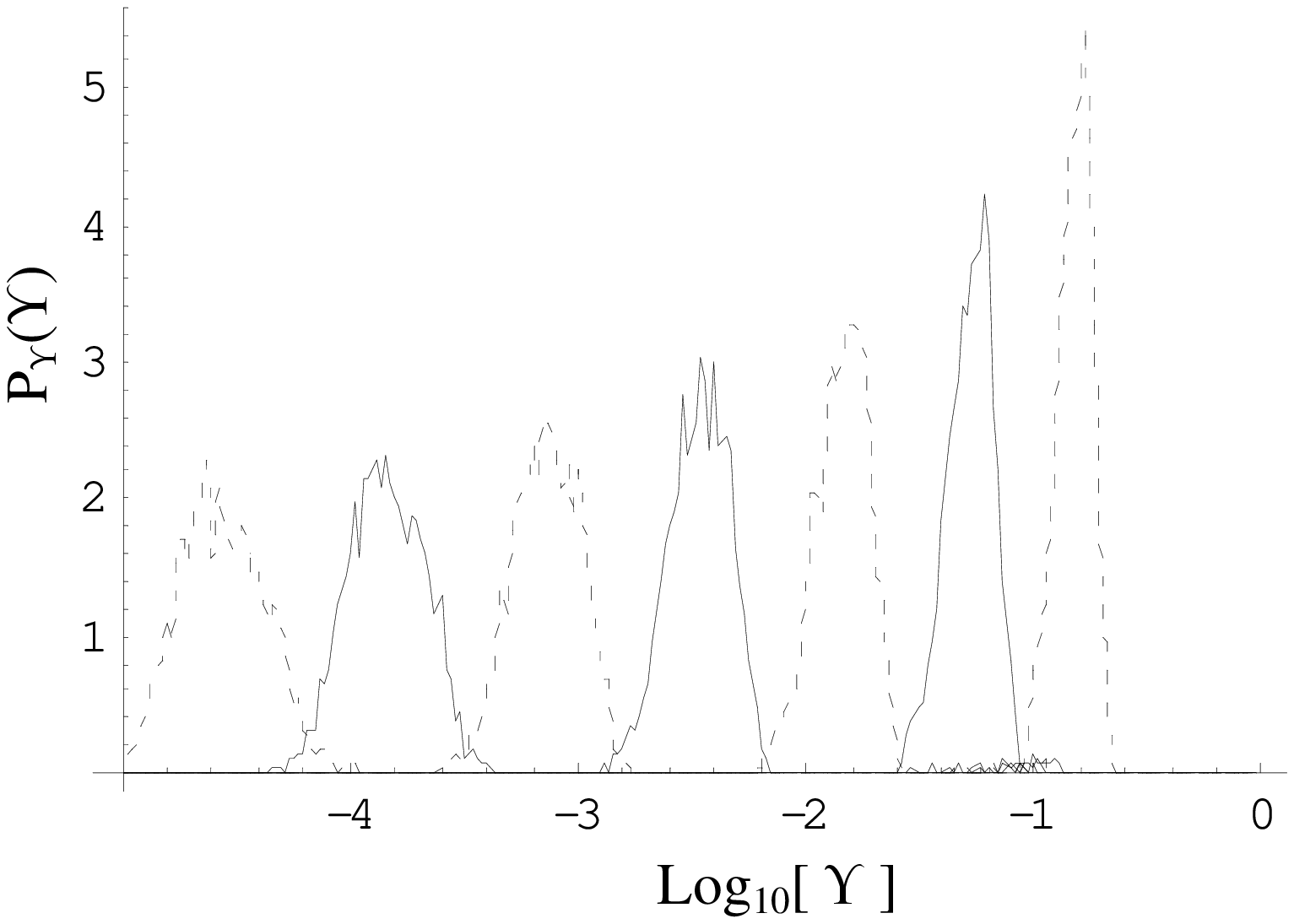}
\includegraphics[angle=0,width=.45\textwidth]{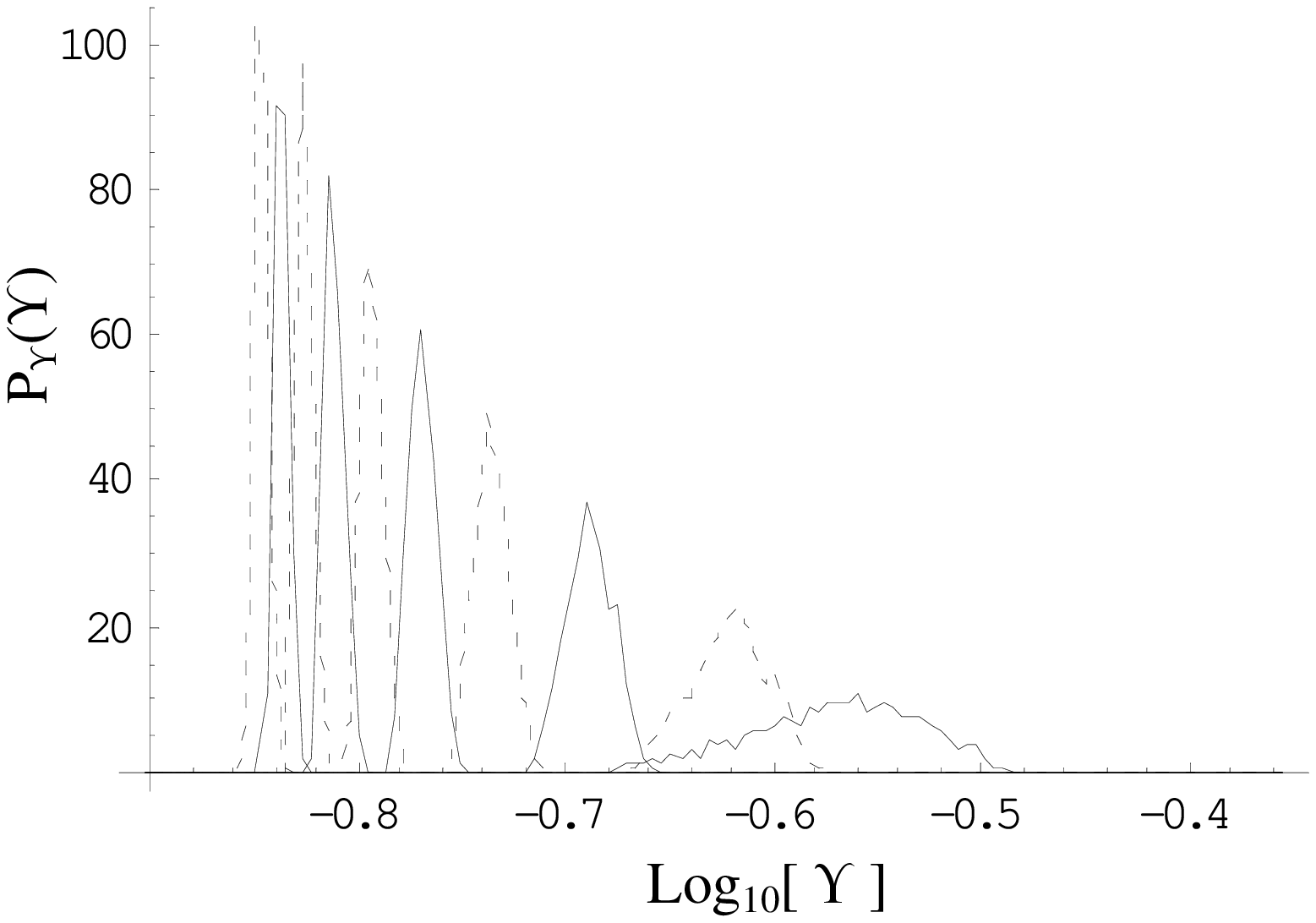}
\caption{\label{prob_ups_s06} Distribution of values of $\Upsilon(s,t,\theta_{j})$ for fixed value $s=0.6$ and several fixed values of $t$.  (Top) From left to right, $t=8$ (dashed), $7$ (solid), $6$ (dashed), $5$ (solid), $4$ (dashed), $3$ (solid), and $2$ (dashed).  (Bottom) From left to right, $t=1/10$ (dashed), $1/9$ (solid),  $1/8$ (dashed), $1/7$ (solid), $1/6$ (dashed), $1/5$ (solid), $1/4$ (dashed), $1/3$ (solid), $1/2$ (dashed), and $1$ (solid).}
\end{figure}

Second, the values of $\Upsilon$ have been calculated according to Eq.~(\ref{Ups1}) for the data obtained in the MSTE.  The conditional PDF $P_{\Upsilon}(\Upsilon \mid s,t)$ was calculated for various fixed values of $s$ and $t$ and these are presented in Figs.~(\ref{prob_ups_s0}) and~(\ref{prob_ups_s06}) for $s=0$ and $s=0.6$, respectively.
For some values of $s$ and $t$, the ratio $\Upsilon^{\text{(max)}}/\overline{\Upsilon}$ is as high as 3 (or greater) and $\Upsilon^{\text{(min)}}/\overline{\Upsilon}$ is as small as 1/3 (or lower).  This means that some grain configurations $\{s_{i}, t_{i}, \theta_{ij}\}$ will occur three times too often or only 1/3 often enough in the MSA ensemble compared to the exact Edwards ensemble.  This effect is most pronounced when $t$ is high and $s$ is low.  However, high values of $t$ are rare to begin with.  Furthermore, the distribution for each pair of values $(s,t)$ is localized with a clear peak and so the majority of grain configurations will have a value of $\Upsilon$ that is \textit{relatively} not very far from $\overline{\Upsilon}$ while being distinctly separate from the $\overline{\Upsilon}$ for other values of $(s,t)$.  These latter considerations imply that the MSA does characterize the organization in the DOS qualitatively, but more effort is needed to show whether it is quantitatively sufficient.

\begin{figure}
\includegraphics[angle=0,width=.45\textwidth]{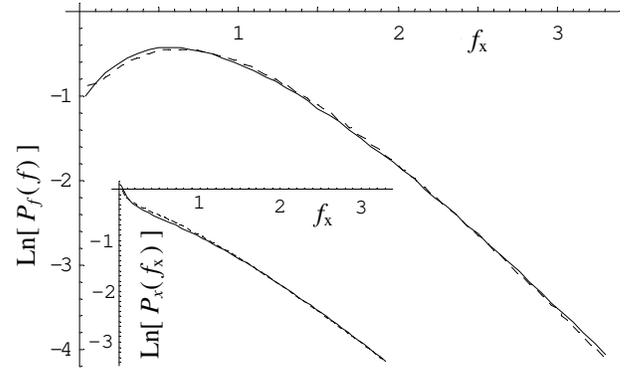}
\caption{\label{compare} Comparison of the curves that were fitted to the empirical $P_{f}(f)$ (large plot) and $P_{x}(f_{x})$ (inset) that resulted from the mean structure transport method using two different sampling methods.  In each plot the solid line uses sampling as Eq.~(\ref{method1}) with quartered fabric, whereas the dashed line used sampling as in Eq.~(\ref{method2}) but with non-quartered fabric.  The results are statistically indistinguishable, lending credence to the mean structure approximation.}
\end{figure}

Third, two different sampling schemes were implemented as presented in Eqs.~(\ref{method1}) and (\ref{method2}).  The results were identical to within the statistical precision of the data, as shown in Fig.~(\ref{compare}).
This shows that the resulting distributions are insensitive to the existence or non-existence of correlations between the Cartesian loads and the contact angles, and this is the essence of the MSA.

\subsection{The Form of the Density of States}

The features of a DOS may be described by two components:  the shape of the accessible regions of the phase space, and the measure that is used within that space.  It is possible that the phase space is not equally accessed by the dynamics of a real packing as it locates and settles into one of the static states.  Perhaps this is more true for the hyperstatic (frictional) case or for other cases less idealized than the one considered here.  The form of the PDFs would then be a reflection of the shape of the measure rather than the shape of the space.  Nevertheless, the use of Edwards' flat measure produced at least the \textit{predominant} features of $P_{F}(F)$, and so those features are attributable to the shape of the space.  The surprising repeatability of $P_{F}(F)$ seen experimentally and in simulations under many conditions and in many non-idealized cases is therefore explained by this fact.

The rise in $P_{F}(F)$ to a peak is not due to a degeneracy of $F$ states in the same way that thermal systems have a distribution shaped by the degeneracy of energy states or momentum magnitudes.  In other words, if a granular contact force $F=\surd[F_{x}^2+F_{y}^2+F_{z}^2]$ had three Cartesian components that were statistically independent, then there would be a volume of phase space $\Omega(F)$ corresponding to each value of $F$ such that $\Omega \to 0$ as $F \to 0$.  If that were the case, then the rise in $P_{F}(F)$ would necessarily begin at the origin.  However, since that is contrary to empirical observations, this sort of Cartesian component degeneracy must not be a dominant feature in the DOS despite the fact that $F$ is a vector magnitude.

An explanation for this begins with the idea that the fundamental unit in a granular packing is not a contact force, but a grain, and so the physics of allowable grain states limits the space (i.e., there must be a grain factor).  For the particular case considered in this paper, there must be six axes in the phase space of single grain states.  The 2D stress tensor has two independent principle stress values, and so at least two of the six axes represent the force state.  These may be $w_{x}$ and $w_{y}$ (or $s$ and $t$).  The other four axes must convey the geometric information, so they may be contact angles.  If the space were given more axes than this then the density of single grain states would be constrained onto a (hyper)surface within that space, but we want the states to fill the volume so that we may examine the behavior of the volume in the limit that one contact force $F_1 \to 0$.  

A fixed value of $F_1$ defines a 5 dimensional region within the single grain space.  Its 5D volume is,
\begin{equation}
\Omega(F_1)=\int\!\!\!\!\int_0^{\infty}\text{d}^2w\ \Omega'(F_1,w_x,w_y)\label{5DOmega}
\end{equation}
where
\begin{eqnarray}
\Omega'(F_1,w_x,w_y)  =  \int\!\!\!\!\int\!\!\!\!\int\!\!\!\!\int \text{d}^4\theta\ \ \Theta(\text{steric exclusion})\nonumber\\*
\ \ \ \times\ \delta\left[F_1-F_1(w_x,w_y,\theta_j)\right]\Theta(F_2)\Theta(F_3)\Theta(F_4)\label{3DOmega}
\end{eqnarray}
is the volume of a 3D hypersurface.  The integrand of this is everywhere nonnegative and for any load state $(w_x>0, w_y>0)$ there exist some angles $\{\theta_{\beta}\}$ such that the integrand is positive.  This is because just three contacts $F_2$, $F_3$, and $F_4$ can support arbitrarily high loads by themselves regardless of the value of $F_1$.  Therefore
\begin{equation}
\Omega'(F_1,w_x,w_y) >  0 \ \ \forall\ w_x>0,\ w_y>0.
\end{equation}
This fact is demonstrated in 2D frictionless numerical simulations where it is seen that a large fraction of the grains have coordination $Z=3$ and yet support compressive loads in both axes, $w_{x}$ and $w_{y}$.  Because of this, it turns out that in Eq.~(\ref{5DOmega}) the integrals in $w$ diverge and $\Omega$ is infinite for all values of $F_1$.  The conclusion is that stable grains with $F_1 \to 0$ are not confined into a vanishing region of the phase space.  This is in contrast to thermal systems where, for example, $p=\surd[p_x^2+p_y^2+p_z^2]$ can be zero if and only if all its statistically independent components become zero so that $\Omega(p) \to 0$ as $p \to 0$.  

There are two key distinctives of the granular phase space.  First, while contact forces are indeed vectors, the stability requirement for the individual grains is so constraining that the components of the vectors cannot be statistically independent.  Therefore, the DOS cannot be uniform in a space defined by the Cartesian axes.  The degeneracy of vector magnitudes does not automatically force $P_F(0)$ to zero.  Second, even the magnitudes of the contact forces sharing the same grain cannot be statistically independent.  Therefore, the DOS cannot be uniform in a space defined by all the force magnitude axes.  The vanishing volume of the non-tensile quadrant near the origin does not automatically force $P_F(0)$ to zero, either.  Instead forces sharing a grain are correlated in some regions of phase space and anti-correlated in other regions, depending upon the the $\{\theta_{\beta}\}$ axes.  It is the existence of anticorrelation that provides the grains no fewer degrees of freedom at $F_1=0$ than they have at any other value of $F_1$.  This will be explained further, below.

This observation about the phase space is the answer why $P_{F}(0)>0$.  Edwards' flat measure predicts it, indicating that the vast majority of metastable packings contain a finite fraction of grains with one or more contacts arbitrarily close to zero force.  In fact, we know this is correct because every time the stress state of a packing is perturbed there is a finite probability that a measurable fraction of the grains will tip and rearrange.  If something in the physics had made the region near zero force to be a vanishing fraction of the accessible space, then a flat measure in the space would have made tipping and rearranging prohibitively improbable.  

Since the volume of phase space does not vanish as $F \to 0$, then what causes the slope in $P_F(F)$ in the region of weak forces?  The answer is that even though $\Omega'$ is nonvanishing as $F_1 \to 0$, it does get somewhat smaller in that limit.  This is because contacts on opposite hemispheres of the grain---say, $F_1$ and $F_3$---are highly correlated.  When $0<F_1 < \left<F\right>$, then $F_3<0$ over a larger region of $\{\theta_{\beta}\}$ than it is when $F_1>\left<F\right>$.  This was proven analytically for a special case in Ref.~\cite{metzger}.  Note that Eq.~\ref{3DOmega} assumes isotropy in the integrand.  Weighting the integrand anisotropically may provide sufficient generality to produce either rising or falling slopes in $P_F(F)$ in the region of weak forces, and this may explain its evolution under slow shearing \cite{antony}.

The reason the Simplest Model of Edwards and Grinev \cite{edwards2} predicts $P_{F}(0)=0$ is because it treats all the input forces and angle cosines $\lambda^2$ as statistically independent.  This implies a phase space $\left\{F_i,\lambda_i^2\right\}$ with many more degrees of freedom than a static grain actually possesses.  Then, the non-negative domains of all the angle cosines ensure that every $F_i$ is positively correlated with $F$, where $F=\lambda_1^2 F_1+\ldots+\lambda_{(Z-1)}^2 F_{(Z-1)}$.  The only way that $F$ can be zero is for all $(Z-1)$ quantities $(\lambda_i^2 F_i)$ to be simultaneously zero, which is vanishingly improbable due to their statistical independence.  This is in contrast to real grains where the neighboring contacts having less than $\pi/2$ radians of angular separation should be anti-correlated.  That is, one contact can lift the load off of its neighboring contact so that if one contact bears more load then the neighbor must bear less.  This anticorrelation allows the grain to be stable with $F =0$ while the other contacts have nonzero forces on them.  That is, the grain finds more ways to be stable with zero force on one of its contacts than simply by having zero force on all of its other contacts.  Without addressing the statistical independence of the inputs, the model could therefore be improved (at the loss of solvability, perhaps) by extending the ranges of $(\lambda_i^2)$ to include some portion of $(\lambda^2) < 0$.  This will account for the range of anti-correlation between neighboring contacts.  Extending the space this way will ensure that $\Omega(F)$ is nonvanishing for $F \to 0$ and will result in $P_F(0)>0$.  This was demonstrated in the numerical solution of the MSTE.  When the values of $(\lambda^2)$ were extracted from the numerical data, it was found that they had a range $-0.5 < \lambda^2 < 1$.  The lower limit reflects steric exclusion, and the upper limit reflects maximal separation on opposite sides of a grain.

\section{Summary and Conclusions}

The use of the FSA makes it possible to solve the DOS based upon Edwards' flat measure in a frictionless granular packing of smooth, round, rigid grains with localized isostacy.  This produces a transport equation that can be solved (at least numerically).  Solution of this transport equation in the MSA was shown to produce the correct features for the contact force distributions.

This success tends to validate Edwards' hypothesis:  the DOS appears to be dominated by features inherent to the static phase space, depending solely upon the packing's present fabric and the stress tensor.  That is, the DOS may not be shaped too significantly during the physics of the dynamic regime before the packings achieve static equilibrium.

The need for further work is apparent. First, the two approximations have not been adequately validated.  The quantitative results should be compared with simulations of rigid, frictionless grains with carefully controlled stress states and carefully measured fabric.  This has not yet been performed because most studies have either included gravity or not reported the stress state or fabric.

Second, solution of the transport equation without the MSA is being developed.  Those results compared against the present study will be an important test of the MSA.

Third, the analysis should be extended and numerical results presented for more general cases.  The case with anisotropic stresses and fabric should demonstrate the qualitative evolution of the PDFs under shearing.  This work has begun, and the initial results are hopeful.

Fourth, the forms of the functions that fit the numerical data for $P_{st}(s,t)$ are tantalizingly simple.  If the cause of this can be identified then a completely analytical solution to the MSTE may be possible.

\begin{acknowledgments}
I am grateful for helpful discussions with Aniket Bhattacharya of the University of Central Florida Physics Department and with Robert Youngquist of NASA's John F. Kennedy Space Center.
\end{acknowledgments}


\begin{thebibliography}{99}
\bibitem{boundary} More importantly, it should be noted that $P_{f}(f)$ along the boundary of a container is really $P_{w}(w)$, and due to the unusual contact geometries in the boundary layer the form of $P_{f}(f)$ is not necessarily the same as it is in the bulk.  These relationships were studied in Refs. \cite{snoeih1} and \cite{snoeih2}, showing at least qualitative if not quantitative similarity between the bulk and boundary distributions.  The numerical simulation studies have published values for the exponential tail decay constant allowing for a more limited quantitative comparison in the bulk.  These simulations have not generally been for the case of isotropic fabric and stress nor always using frictionless, rigid grains, so comparison with the bulk in the region of weak forces can at the present be only qualitative.
\end{thebibliography}
\end{document}